%% file: book.tex
\documentclass[graybox,envcountchap,sectrefs]{svmono}

\usepackage{type1cm}         

\usepackage{makeidx}         
\usepackage{graphicx}        
\usepackage{multicol}        
\usepackage[bottom]{footmisc}

\usepackage{newtxtext}       %
\usepackage[varvw]{newtxmath}       

\usepackage{tikz}
\usetikzlibrary{arrows.meta}
\usetikzlibrary{arrows.meta,calc,angles,quotes}
\usetikzlibrary{positioning,fit,backgrounds,shapes.geometric}
\usepackage{pgfplots}
\pgfplotsset{compat=1.16}

\usepackage{hyperref}

\usepackage{comment}
\usepackage{gensymb}
\usepackage{listings} 
\usepackage{xcolor} 

\usepackage{wrapfig}
\usepackage[T1]{fontenc}
\usepackage{listings}
\usepackage{xcolor}
\usepackage{courier}
\usepackage{caption}

\usepackage{graphicx}
\usepackage{subcaption}

\usepackage{tabularx}
\usepackage{booktabs} 
\usepackage{hyperref}

\makeindex             

\begin{document}

\author{Xiao Wang, Jayasai Rajagopal, Md Safaiat Hossain, Peng Chen, Mohamed Wahib, Enzhi Zhang, Emma J. Reid}
\title{Efficient Computing for Medical Image Acquisition and Reconstruction}
\maketitle

\frontmatter

\tableofcontents

\include{acronym}

\mainmatter
\include{chapter}

\include{acknowledgement}

\backmatter
\printindex

\bibliography{reference.bib} 
\bibliographystyle{spmpsci} 

\end{document}

%% file: acronym.tex
%
%

\extrachap{Acronyms}

\begin{description}[CABR]
\item[AI]{Artificial Intelligence}
\item[CPU]{Central Processing Unit}
\item[CT]{Computed Tomography}
\item[FBP]{Filtered Backprojection}
\item[FFT]{Fast Fourier Transform}
\item[GPU]{Graphics Processing Unit}
\item[HPC]{High Performance Computing}
\item[ICD]{Iterative Coordinate Descent}
\item[MAP]{Maximum A Posteriori}
\item[MBIR]{Model-Based Iterative Reconstruction}
\item[MLEM]{Maximum-Likelihood Expectation-Maximization}
\item[MRI]{Magnetic Resonance Imaging}
\item[NUFFT]{Nonuniform Fast Fourier Transform}
\item[OSEM]{Ordered-Subsets Expectation-Maximization}
\item[PET]{Positron Emission Tomography}
\item[SENSE]{Sensitivity Encoding}
\item[SPECT]{Single-Photon Emission Computed Tomography}
\end{description}

%% file: chapter.tex
%
%
%
\chapter{\Large{Efficient Computing for Medical Image Acquisition and Reconstruction}}
\label{intro} 

\abstract{
Medical imaging systems such as Computed Tomography (CT), Magnetic Resonance Imaging (MRI), Positron Emission Tomography (PET), and Single-Photon Emission Computed Tomography (SPECT) do not directly acquire images. Instead, they measure physical signals that encode anatomical or physiological information, and image reconstruction recovers the underlying image by solving an inverse problem. Although these imaging modalities are governed by different imaging physics, they share a common computational framework that naturally connects medical physics, linear algebra, probability, numerical optimization, and efficient computing.\\
As medical imaging systems acquire increasingly large and higher-dimensional datasets, image reconstruction has become one of the primary computational bottlenecks in modern medical imaging. Advanced reconstruction methods, including analytical reconstruction, iterative optimization, and statistical model-based reconstruction, substantially improve image quality while reducing radiation dose or scan time, but at significantly increased computational cost. Efficient computing has therefore become essential for achieving clinically practical reconstruction times.\\
This chapter presents a unified computational perspective on medical image acquisition and reconstruction across CT, MRI, PET, and SPECT. It first reviews the imaging physics and data acquisition process for each modality and derives a generalized mathematical framework for image reconstruction. Building on this framework, the chapter discusses analytical, iterative, and statistical reconstruction methods together with their computational characteristics. Finally, it examines efficient computing considerations, including optimization algorithms, physics-aware forward operators, memory-efficient implementations, and parallel computing strategies. Together, these topics demonstrate how the integration of imaging physics, mathematical modeling, and efficient computing enables accurate and scalable medical image reconstruction.
}

\section{Introduction to Efficient Computing}
\label{sec:intro-to-efficient-computing}
Medical imaging has become one of the most computationally intensive fields in healthcare. Modern imaging systems routinely acquire hundreds of millions of measurements from detector sensors with increasingly higher spatial resolution, temporal resolution, and dimensionality. Transforming these measurements into clinically useful images requires sophisticated reconstruction algorithms involving signal processing, linear algebra, statistical inference, numerical optimization, and, more recently, Artificial Intelligence (AI). Consequently, modern medical imaging systems must process enormous amounts of data while operating under strict constraints on reconstruction time, memory capacity, energy consumption, hardware cost, and clinical workflow.

Throughout this chapter, we adopt a broad view of \emph{efficient computing}. Rather than referring only to fast hardware or parallel computing, efficient computing encompasses the design of mathematical models, reconstruction algorithms, numerical optimization methods, software implementations, and computing architectures that maximize reconstruction quality while minimizing computational cost. From this perspective, efficient computing spans the entire imaging pipeline, including both GPU acceleration and distributed computing. High-Performance Computing (HPC) is therefore one important component of efficient computing rather than its sole focus. In fact, a central theme of this chapter is that modern medical imaging cannot be fully understood from the perspective of medical physics, signal processing, or computer science alone. Instead, it arises from the interaction of three disciplines: a) physics determines how measurements are acquired, b) mathematics formulates the underlying physical processes, and c) efficient computing determines whether these mathematical models can be solved within practical clinical constraints.

Unlike conventional photography, most medical imaging systems do not directly capture images. Instead, they first acquire indirect physical measurements using specialized detector systems and subsequently recover images computationally through a process known as \emph{image reconstruction}. For example, Computed Tomography (CT) reconstructs images from X-ray projection measurements, Magnetic Resonance Imaging (MRI) reconstructs images from measurements acquired in k-space, and Positron Emission Tomography (PET) and Single-Photon Emission Computed Tomography (SPECT) reconstruct images from detected gamma-ray events. Recovering an image from these indirect measurements is fundamentally an \emph{inverse problem}, which means that the unknown image must be inferred from a limited set of indirect measurements rather than observed directly~\cite{fessler2008image}. Therefore, image reconstruction lies at the heart of modern medical imaging and has become one of its most computationally demanding components.

The computational complexity of image reconstruction arises from three interacting factors. First, increasingly accurate imaging physics requires increasingly sophisticated forward operators that more faithfully describe detector geometry, system response, and image formation. Second, modern reconstruction algorithms, including iterative reconstruction, statistical reconstruction, Bayesian inference, compressed sensing, and AI-assisted methods, repeatedly solve large-scale optimization problems that require many computational iterations. Third, advances in imaging hardware continue to generate increasingly large multidimensional datasets that must be processed efficiently. Together, these trends have made computation, memory bandwidth, and data movement major performance bottlenecks in modern imaging systems.

Efficient computing addresses these challenges at multiple levels. At the algorithmic level, \emph{analytical reconstruction} methods exploit mathematical transforms such as the Radon transform and the Fourier transform to obtain computationally efficient closed-form solutions. \emph{Iterative and statistical iterative reconstruction} methods improve image quality by employing more accurate forward operators, statistical noise models, and prior information, albeit at substantially greater computational cost. At the implementation level, optimized numerical algorithms, memory-efficient data structures, reduced-precision arithmetic, and hardware-aware software design further reduce execution time and memory consumption. Finally, hardware acceleration through multicore central Processing Units (CPUs), Graphics Processing Units (GPUs), and distributed-memory systems enables computationally intensive reconstruction algorithms to achieve clinically practical reconstruction times.

The distinction between efficient computing and HPC can be understood from the life cycle of a reconstruction algorithm. During algorithm development, researchers often prioritize reconstruction accuracy and may employ HPC systems to train AI models, evaluate sophisticated reconstruction methods, and perform large-scale numerical experiments. During clinical deployment, however, the priorities shift toward reconstruction latency, memory usage, hardware cost, robustness, and integration into routine clinical workflows. Algorithms developed using HPC resources must therefore ultimately be translated into efficient implementations suitable for imaging scanners and clinical workstations. In this sense, HPC accelerates research and algorithm development, whereas efficient computing enables practical clinical deployment.

This chapter develops a unified computational perspective on medical image acquisition and reconstruction across CT, MRI, PET, and SPECT. Rather than treating each modality independently, it demonstrates how they can all be formulated within a common imaging framework while differing primarily in their imaging physics, forward operators, statistical assumptions, and optimization algorithms. Building on this unified perspective, the chapter examines the computational challenges that arise across imaging modalities and discusses efficient computing techniques that enable high-quality image reconstruction in modern medical imaging systems.

\section{Computing for Medical Imaging: A Pipeline View}
\label{sec:pipeline}

\begin{figure}[b]
\centering
\includegraphics[width=\columnwidth]{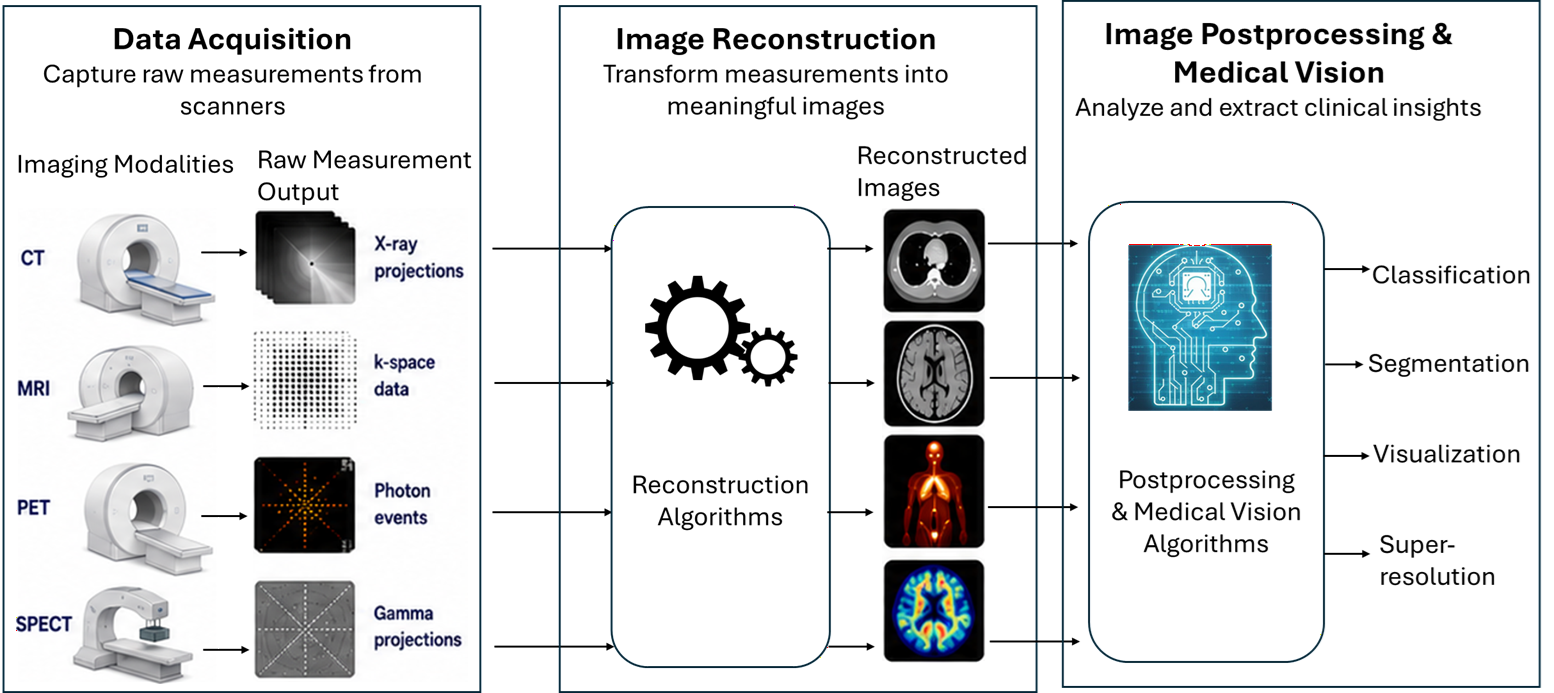}
\caption{Pipeline view of modern medical imaging. Raw measurements acquired from different imaging modalities are reconstructed into medical images and subsequently analyzed through image postprocessing and medical vision. Efficient computing supports every stage of the pipeline, from data acquisition and image reconstruction to downstream image analysis.}
\label{fig:pipeline}
\end{figure}

Modern medical imaging systems are best viewed as computational pipelines rather than isolated imaging devices. Information acquired by the imaging hardware is progressively transformed through a sequence of computational stages, ultimately producing clinically meaningful interpretations. Although the underlying imaging physics differs among CT, MRI, PET, and SPECT, all four modalities follow the same high-level computational workflow consisting of data acquisition, image reconstruction, and image postprocessing, as illustrated in Fig.~\ref{fig:pipeline}. Efficient computing therefore plays a central role throughout the entire imaging pipeline rather than within a single reconstruction algorithm.

The first stage is \emph{data acquisition}. During acquisition, the imaging system measures physical signals produced by the patient using specialized detector hardware. The form of these measurements depends on the imaging modality. CT acquires X-ray projection measurements, MRI samples spatial-frequency measurements in k-space, while PET and SPECT detect emitted gamma-ray photons. These measurements encode information about the patient's anatomy or physiology but do not yet form an image. The computational objective of this stage is to acquire, transfer, and organize large volumes of measurement data accurately and efficiently. Modern imaging systems often generate data at several gigabytes per second, making efficient data movement, streaming, buffering, and storage important computational considerations.

The second stage is \emph{image reconstruction}. Image reconstruction transforms the acquired measurements into clinically interpretable images by solving an inverse problem~\cite{fessler2008image}. As shown later in this chapter, despite their different imaging physics, CT, MRI, PET, and SPECT can all be formulated within a common computational framework, allowing many reconstruction algorithms to be understood from a unified mathematical perspective. This stage forms the computational core of most medical imaging systems and is the primary focus of this chapter. Reconstruction algorithms range from analytical methods based on the Radon transform and Fourier transform to iterative optimization~\cite{kak2001principles}, statistical reconstruction, Bayesian inference, and AI-assisted methods. As reconstruction models become increasingly accurate, their computational complexity also increases substantially~\cite{wang2016high,Wang2017Massively3D}. Consequently, efficient computing is essential for achieving clinically practical reconstruction times while maintaining high image quality.

The final stage is \emph{image postprocessing}.
Unlike image reconstruction, which estimates an image from physical measurements, image postprocessing assumes that the image has already been reconstructed and instead focuses on extracting clinically meaningful information~\cite{dougherty2009digital}. Typical postprocessing tasks include image enhancement, registration, segmentation, classification, detection, and AI-assisted image interpretation. Although postprocessing is an important component of the overall imaging workflow, its computational objectives differ from those of image reconstruction because the input is already an image rather than raw measurement data.

Each stage of the imaging pipeline presents a different computational challenge. Data acquisition emphasizes efficient measurement collection, data transfer, and storage. Image reconstruction focuses on solving large-scale inverse problems through mathematical modeling and numerical computation. Image postprocessing concentrates on extracting clinically meaningful information from reconstructed images. Although these stages solve different computational problems, they are tightly coupled. Improvements in acquisition often increase the demands placed on reconstruction, while advances in reconstruction directly influence the performance of downstream image analysis.

This chapter primarily focuses on the first two stages of the pipeline: data acquisition and image reconstruction. The following sections first review the physical principles of data acquisition for CT, MRI, PET, and SPECT, and then develop a unified mathematical framework for image reconstruction. Building upon this framework, the chapter discusses analytical, iterative, and statistical reconstruction methods together with the efficient computing techniques that enable their practical implementation in modern medical imaging systems.

\section{Data Acquisition}
\label{sec:data acquisition}

\begin{figure}[h]
\centering
\includegraphics[width=.4\textwidth]{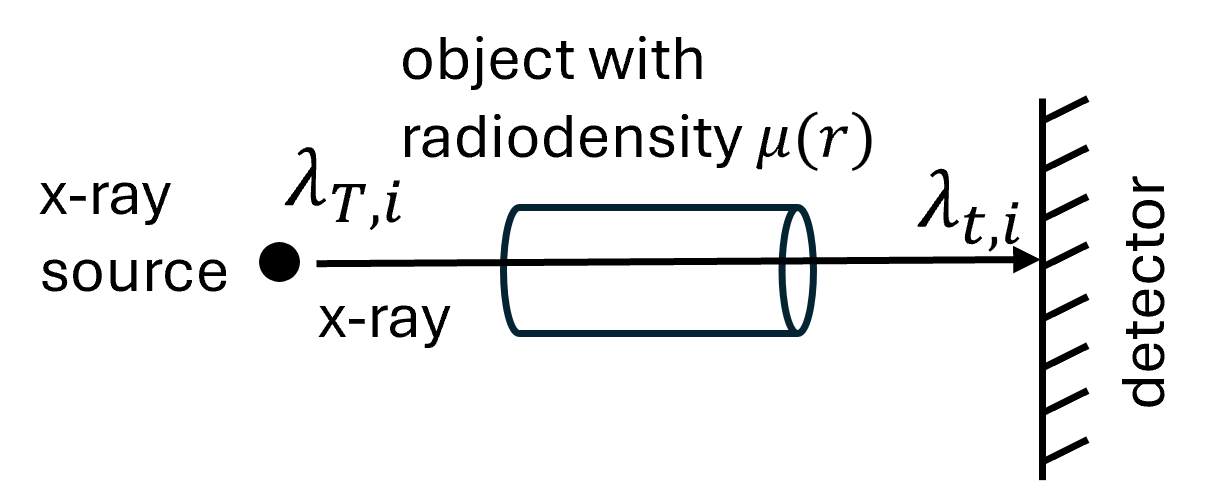}
\caption{Incident X-ray photon counts $\lambda_{T,i}$ from an X-ray beam are attenuated by object. After attenuation, $\lambda_{t,i}$ photon counts are received by an X-ray detector.}
\label{fig:photon_attenuation}
\end{figure}

\subsection{Data Acquisition for X-ray Computed Tomography}
\label{subsec:ct-data-acquisition}
The physical principle underlying CT acquisition is X-ray attenuation. Consider a single X-ray indexed by $i$, the number of X-ray photons emitted from the X-ray source is referred to as \emph{incident photon counts} and is denoted as $\lambda_{T,i}$. As the X-ray passes through an object along the path of the X-ray, photons are absorbed (also known as \emph{attenuation}) or scattered by the object, reducing the number of photons that reach the detector. The remaining photons that are captured by an X-ray detector are referred to as the \emph{transmitted (or received) photon counts}, also denoted as $\lambda_{t,i}$. Figure~\ref{fig:photon_attenuation} illustrates the attenuation effect, showing the incident and received photons in the figure. The \emph{linear attenuation coefficient} for an object represents the object's radiodensity, which measures a material's ability to absorb X-rays, and is commonly represented clinically in Hounsfield units~\cite{hsieh2003computed}. If the linear attenuation coefficient can be described as a distribution function $\mu(\mathbf{r})$, where $\mathbf{r}$ represents 3D spatial coordinates, then the relationship among $\lambda_{T,i}$, $\lambda_{t,i}$ and $\mu(\mathbf{r})$ can be described by the \emph{Beer--Lambert law}~\cite{swinehart1962beer, kak2001principles} as follows:

\begin{equation}
\lambda_{t,i} = \lambda_{T,i} e^{-y_i} =  \lambda_{T,i} e^{-\int \mu(\mathbf{r}) \, \mathrm{d}\mathbf{r}} \ ,
\label{eq:beer_lambert}
\end{equation}
where $y_i$ is the measurement associated with the $i^{th}$ X-ray path and represents the line integral of the object's radiodensity along the path, namely $y_i =\int \mu(\mathbf{r}) \, \mathrm{d}\mathbf{r}$.
From the above equation, we can also derive that $y_i$ can be computed as:
\begin{equation}
y_i = -\log\left(\frac{\lambda_{t,i}}{\lambda_{T,i}}\right) \ ,
\label{eq:X-ray-projection}
\end{equation}
where taking the negative logarithm converts the exponential attenuation model into a linear line integral.

\begin{figure}[h]
\centering
\includegraphics[width=.9\textwidth]{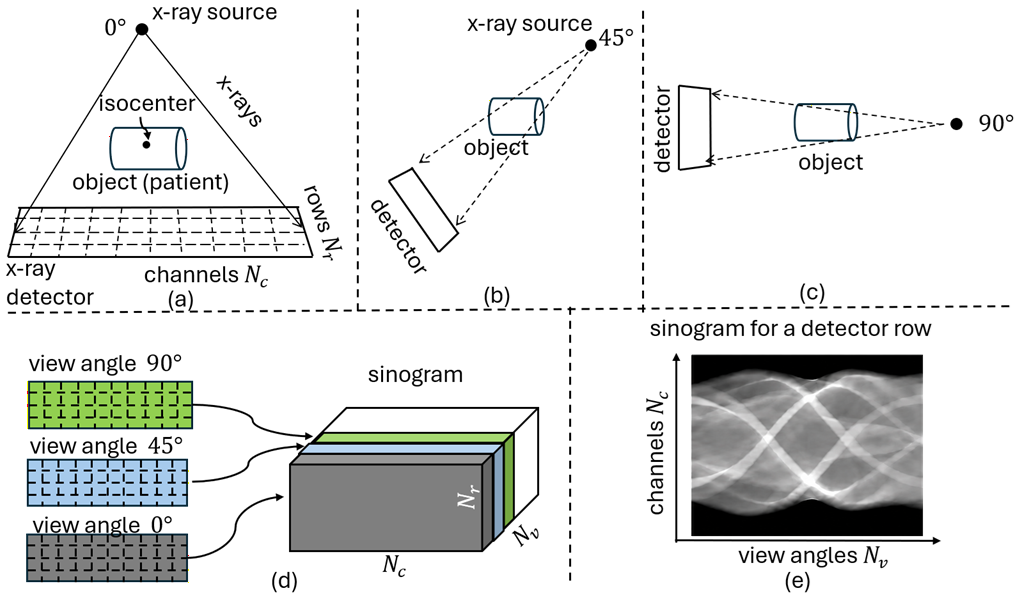}
\caption{(a) Illustration of CT data acquisition setup at view angle 0\degree. (b) and (c) The X-ray source and detector rotate around the patient to 45\degree and 90\degree view angles.  (d) The acquired projections from all view angles are stacked up to form a three-dimensional sinogram $\mathbf{Y}$ of size $N_v \times N_c \times N_r$, where $N_v$, $N_c$, and $N_r$ denote the numbers of projection view angles, detector channels, and detector rows, respectively. (e) For a fixed detector row, the sinogram takes the shape of sinusoidal traces.}
\label{fig:ct_acquisition}
\end{figure}

During CT acquisition, an X-ray source and detector are mounted on opposite ends of a rotating gantry that revolves around the patient. Figures~\ref{fig:ct_acquisition}(a) (b) (c) show the CT scan rotation at view angles 0\degree, 45\degree, and 90\degree. The patient is positioned near the center of rotation, known as the \emph{isocenter}. When the scanner rotates to each view angle, the detector records the transmitted photon counts from all detector elements. Applying the negative logarithm in Equation~(\ref{eq:X-ray-projection}) to measurements from all detector elements produces a two-dimensional data matrix at each view angle, known as a \emph{projection}.

A modern X-ray detector for CT is organized into smaller sensor arrays, consisting of detector \emph{channels}, denoted as $N_c$, and detector \emph{rows}, denoted as $N_r$~\cite{wang2021physics,kalender2006x}. Detector channels sample the transverse direction of the X-ray beam, whereas detector rows are aligned along the patient's head-to-foot direction. Consequently, a projection from each view angle is organized into a two-dimensional array of size $N_c \times N_r$. As the gantry rotates, this process of acquiring projections is repeated over many view angles, denoted as $N_v$, and each view angle produces different projections of the same patient. 

After acquiring projections from all the view angles, the projections are stacked along the view angles to form the complete CT measurement dataset, as illustrated in Figure~\ref{fig:ct_acquisition}(d). This complete measurement dataset, denoted as $\mathbf{Y}$ in this chapter, is referred to as the three-dimensional \emph{sinogram} and it has a dimension size of $N_c \times N_r \times N_v$. For a fixed detector row, the corresponding two-dimensional slice of the sinogram $\mathbf{Y}$ exhibits characteristic sinusoidal traces, as shown in Figure~\ref{fig:ct_acquisition}(e). This sinusoidal pattern gives the sinogram its name. Note that the sinogram contains all measurements, but it bears little visual resemblance to the patient's anatomy and cannot be directly interpreted clinically. Therefore, recovering clinically interpretable images from a sinogram requires image reconstruction, which will be explained in later sections.

Advances in detector technology have substantially increased the size of modern CT acquisition datasets~\cite{hsieh2003computed}. The acquired projections are often collected over a large number of view angles, with each view containing hundreds to thousands of detector channels $N_c$ and many detector rows $N_r$. The scale of these dimensions directly determines the amount of acquired data and significantly influences storage requirements, memory traffic, and the computational complexity of reconstruction algorithms.
Representative hardware specifications of several commercial CT systems are summarized in Table~\ref{tab:commercial_ct_systems}. The reconstructed image volume is usually generated on a $512 \times 512$ image grid for each slice, although higher spatial resolutions are increasingly available.

\begin{table}[ht]
\centering
\caption{Representative medical hardware specifications of modern CT systems. The listed values are typical configurations reported in vendor documentation.}
\label{tab:commercial_ct_systems}

\resizebox{\linewidth}{!}{
    \begin{tabular}{llccc}
    \hline
    \textbf{Vendor} &
    \textbf{CT System} &
    \textbf{Rows ($N_r$)} &
    \textbf{Channels ($N_c$)} &
    \textbf{Views ($N_v$)} \\
    \hline
    
    \href{https://www.gehealthcare.com}{GE HealthCare} &
    Discovery CT750 HD &
    64 &
    888--984 &
    $\approx$1000 \\
    
    \href{https://www.siemens-healthineers.com}{Siemens Healthineers} &
    SOMATOM Definition Flash &
    64 &
    $\approx$800--1000 &
    $\approx$1150 \\
    
    \href{https://www.siemens-healthineers.com}{Siemens Healthineers} &
    SOMATOM Force &
    192 &
    $>$900 &
    $\approx$1200 \\
    
    \href{https://global.medical.canon}{Canon Medical Systems} &
    Aquilion ONE &
    320 &
    896 &
    $\approx$1000 \\
    
    \href{https://www.philips.com/healthcare}{Philips Healthcare} &
    Brilliance iCT &
    128 &
    $\approx$800--1000 &
    $\approx$1000 \\
    
    \href{https://www.united-imaging.com}{United Imaging Healthcare} &
    uCT 960+ &
    160 &
    $\approx$900 &
    $\approx$1200 \\
    
    \hline
    \end{tabular}
}

\end{table}

\begin{figure}[h]
\centering
\includegraphics[width=\textwidth]{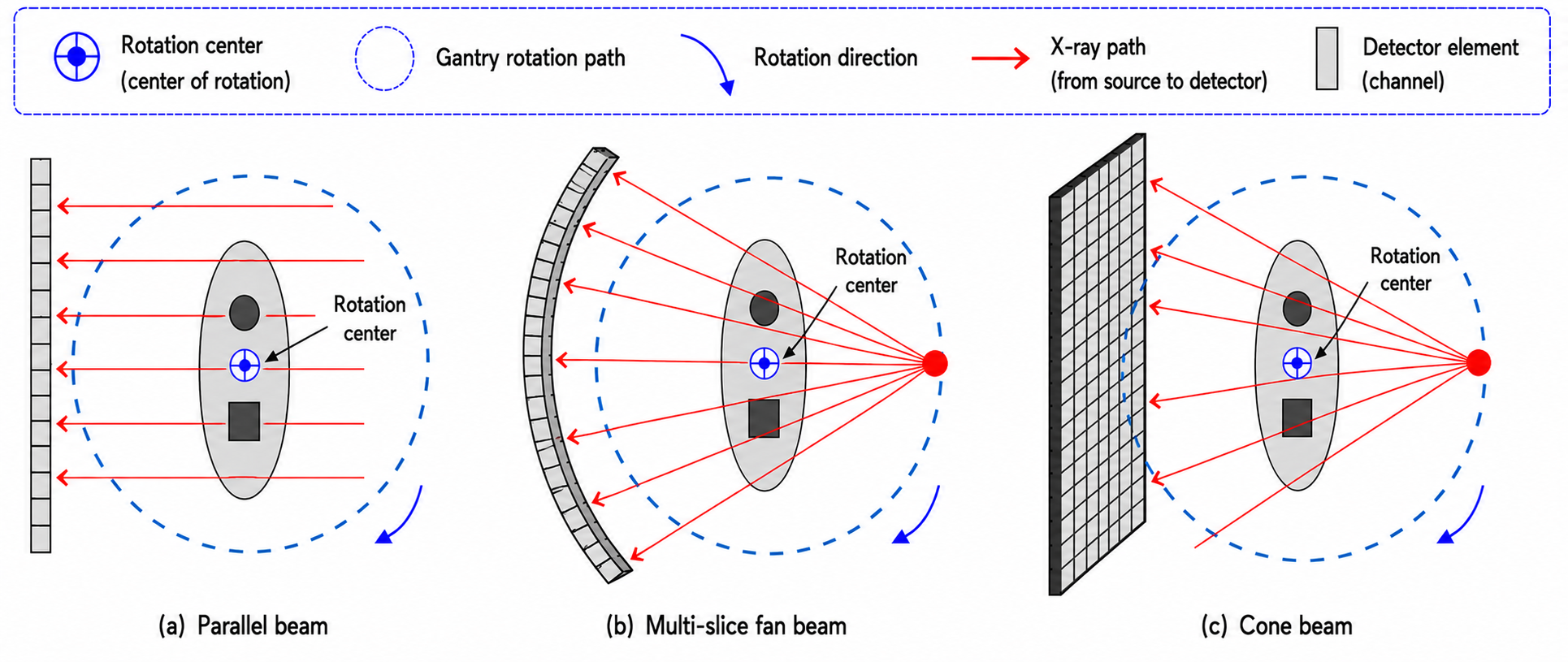}
\caption{Common CT acquisition geometries. (a) Parallel-beam geometry using parallel X-rays and a one-dimensional detector array. (b) Multi-slice fan-beam geometry using a point X-ray source and a two-dimensional arc detector array. (c) Cone-beam geometry using a point X-ray source and a two-dimensional flat-panel detector for volumetric imaging.}
\label{fig:X-ray.shapes}
\end{figure}
Several acquisition geometries have been developed for CT imaging, as illustrated in Figure~\ref{fig:X-ray.shapes}. The three principal geometries are parallel-beam, multi-slice fan-beam, and cone-beam CT, each representing a different configuration of the X-ray source and detector. Regardless of the acquisition geometry, the resulting sinogram serves as the input to the image reconstruction algorithms discussed in Sec.~\ref{sec:image reconstruction}.
In the parallel-beam geometry shown in Figure~\ref{fig:X-ray.shapes}(a), all X-ray rays are parallel to one another. This geometry leads to a mathematically simple acquisition model and forms the basis of many theoretical developments in tomographic reconstruction. Although parallel-beam CT was used in early first-generation CT scanners, it has largely disappeared from clinical CT because of its low acquisition efficiency. It remains widely used in scientific imaging applications for synchrotron, neutron tomography, and X-ray microscopy, where parallel beam illumination is naturally available.

In the multi-slice fan-beam geometry shown in Figure~\ref{fig:X-ray.shapes}(b), a point X-ray source emits a diverging fan-shaped beam that is measured by a curved two-dimensional detector array. Fan-beam geometry makes much more efficient use of X-ray photons and therefore became the standard configuration for clinical CT. During clinical scans, the patient table typically moves continuously through the rotating gantry, producing a helical (spiral) scan trajectory~\cite{kalender2006x,wang2021physics}, which enables rapid volumetric imaging with improved patient throughput.

The cone-beam geometry shown in Figure~\ref{fig:X-ray.shapes}(c) replaces the arc detector with a two-dimensional flat-panel detector, allowing an entire volume to be acquired in a single rotation. Cone-beam CT is widely used in dental imaging, maxillofacial imaging, image-guided surgical interventions, and breast tomosynthesis, where its large volumetric coverage and compact hardware provide significant advantages.

\subsection{Data Acquisition for MRI} \label{sec:MRIdata}

\begin{figure}[t]
    \centering
    \includegraphics[width=.8\linewidth]{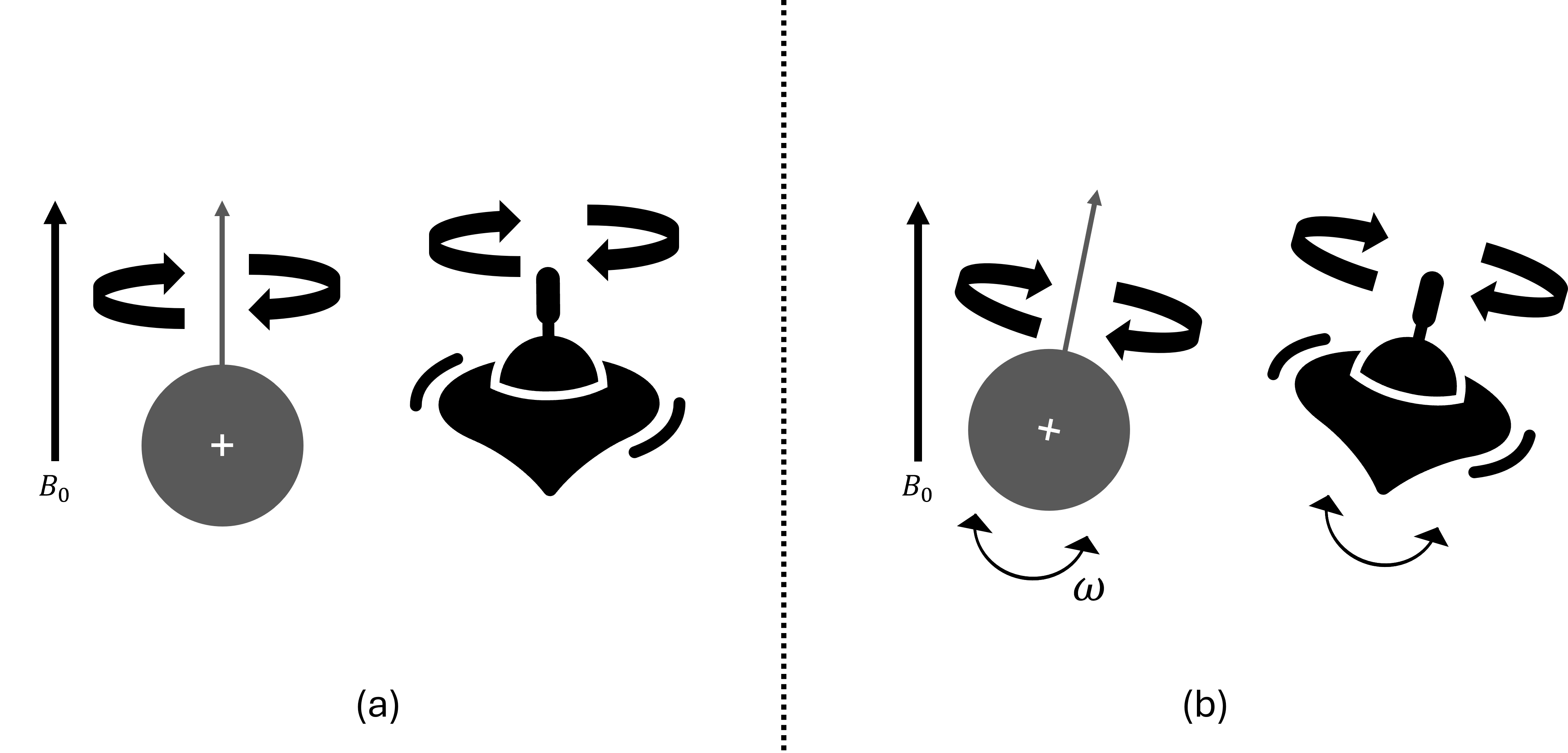}
    \caption{Illustration of proton spin (a) and Larmor precession (b). In the presence of a magnetic field $B_0$, the proton spin aligns with the magnetic field. The proton precesses around the axis of rotation at the Larmor frequency ($\omega$). The precession of a proton's magnetic moment about the main magnetic field $B_0$ can be understood by analogy to the wobbling motion of a spinning top.}
    \label{fig:MR-protons}
\end{figure}

While CT reconstructs the linear attenuation coefficient, $\mu(\mathbf{r})$, which characterizes the radiodensity of each voxel, MRI reconstructs tissue properties from electromagnetic signals, $m(\mathbf{r})$, generated by hydrogen nuclei (protons) inside the patient. These signals arise from the interaction between hydrogen nuclei and externally applied magnetic fields. Because different biological tissues contain different concentrations of hydrogen and exhibit different magnetic relaxation properties, the measured signals provide excellent soft-tissue contrast. Consequently, MRI has become one of the most important imaging modalities for neurological, musculoskeletal, cardiovascular, and abdominal imaging.

\begin{figure}[t]
    \centering
    \includegraphics[width=\linewidth]{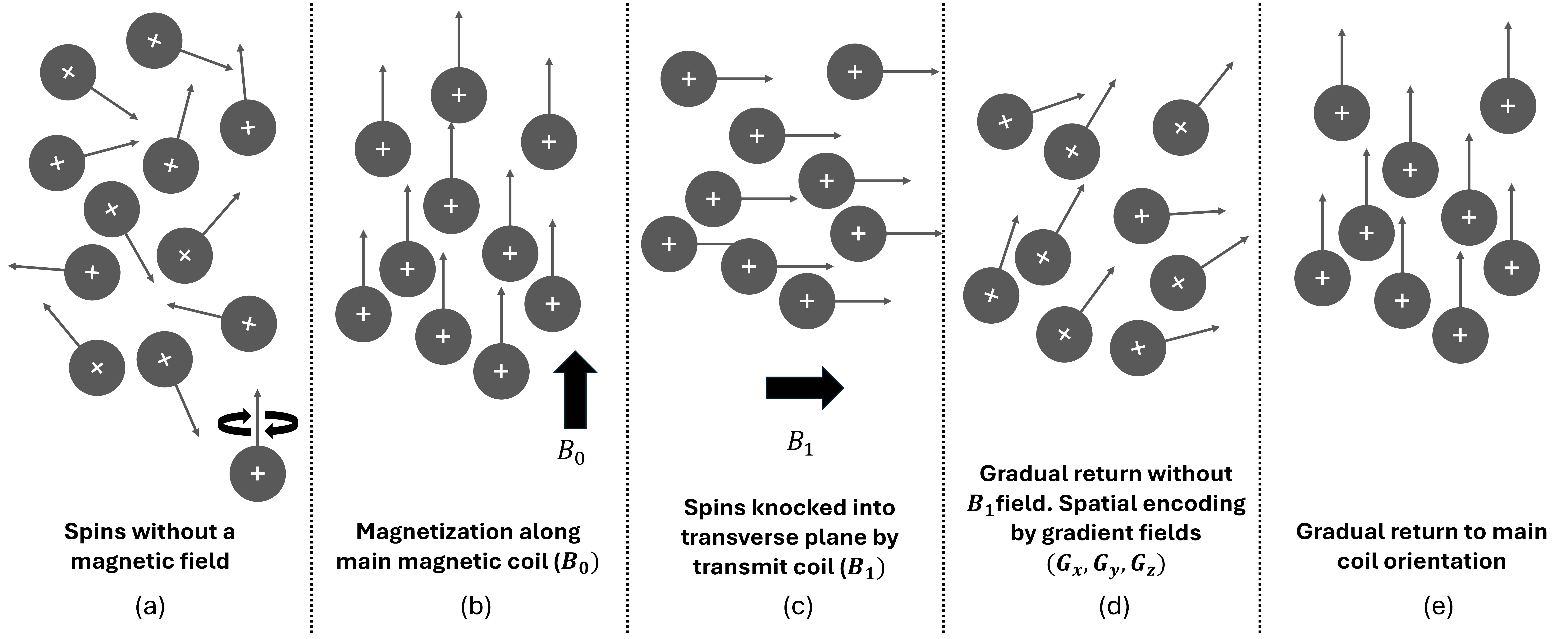}
    \caption{Illustrates different steps in MR signal acquisition. (a) Without a magnetic field, protons spin in random directions. (b) The activation of the main magnetic field, $B_0$, aligns proton spins with the magnetic field. (c)  Oscillating magnetic field $B_1$ from transmit coils knocks spins into the transverse plane. (d) When the $B_1$ field is removed, the protons gradually return to their alignment with $B_0$, but at different rates for each proton. The gradient coils are used to encode positional information. (e) Protons are finally returned to their alignment with $B_0$.}
    \label{fig:MR-acquisition}
\end{figure}

The physical principle underlying MRI is the magnetic behavior of hydrogen protons. Hydrogen atoms are particularly suitable for MRI because they are abundant throughout the human body, primarily in water and fat, and possess a strong magnetic response. Each hydrogen proton has an intrinsic quantum mechanical property known as \emph{spin}, causing it to behave like a microscopic magnetic dipole. 
When the patient is placed inside the MRI scanner, a strong static magnetic field, denoted by $B_0$, aligns these microscopic magnetic dipoles predominantly along the direction of the magnetic field, producing a small net magnetization. In addition to aligning with $B_0$, the magnetization continuously rotates about the rotational axis aligned with the field direction. This rotation is known as \emph{Larmor precession}, and its angular frequency is:
\begin{equation}
\omega=\gamma B_0 \ , 
\end{equation}
where $\omega$ denotes the angular precession frequency, $\gamma$ is the gyromagnetic ratio of the nucleus, and $B_0$ is the strength of the main magnetic field. The precession frequency increases linearly with magnetic field strength \cite{PlewesMRIPrimer}. A visualization of spin and precession is depicted in Figure~\ref{fig:MR-protons}. In panel (a), the spin of a proton aligns with the $B_0$ magnetic field. In panel (b), the proton precesses around the central axis of rotation at the Larmor frequency ($\omega$). Both spin and precession properties can be understood by analogy to the motion of a spinning top, which wobbles around a central spin axis.

The aligned magnetization does not produce a measurable MRI signal by itself. To generate a signal, the scanner applies a short radio-frequency (RF) pulse using a \emph{transmit RF coil}. This oscillating magnetic field, denoted by $B_1$, is applied at the Larmor frequency and tips the net magnetization away from the direction of $B_0$ into the transverse plane. Immediately after the RF pulse is turned off, the transverse magnetization continues to precess while gradually returning to equilibrium through the tissue-dependent relaxation processes characterized by the \emph{$T_1$ and $T_2$ relaxation times}. $T_1$ relaxation measures the time for longitudinal magnetization to return towards equilibrium and generally scales with $B_0$. $T_2$ relaxation is a measure of the phase decoherence of protons after the removal of the $B_1$ field.

The rotating transverse magnetization produces a time-varying magnetic field that induces an electrical voltage in the \emph{receiver coil} according to Faraday's law of electromagnetic induction. This induced voltage from the receiver coil forms the raw MRI signal measured by the scanner. If all protons precessed at the same frequency, the receiver coil would detect only a single combined signal with no information about where the signal originated inside the patient. Consequently, additional magnetic fields are required to encode spatial information \cite{PlewesMRIPrimer}.


Spatial encoding is achieved using three \emph{gradient coils} for three orthogonal spatial directions $x-y-z$ that generate linearly varying magnetic fields superimposed on the main magnetic field $B_0$. Unlike the uniform field $B_0$, gradient fields vary linearly with spatial position. For example, an $x$-gradient produces the magnetic field:
$
B(x)=B_0+G_x x,
$
where $G_x$ denotes the gradient strength. Since the Larmor frequency is proportional to the magnetic field, protons at different spatial positions now precess at slightly different frequencies and accumulate different phases. By appropriately varying the gradient waveforms, every spatial location produces a unique frequency and phase evolution,
thereby encoding spatial information into the measured signal. Table~\ref{tab:mri_coils} summarizes the major magnetic fields and coils involved in the MRI acquisition process. Figure~\ref{fig:MR-acquisition} displays the signal acquisition process and the role of each magnetic coil. Before any magnetization, proton spins face in random directions in (a). After the static magnetic field ($B_0$) is turned on, the spins align to be parallel to that field in (b). When the RF pulse ($B_1$) is turned on, the spins fall into the transverse plane and perpendicular to the $B_0$ field in (c). After the $B_1$ field is removed, the protons begin to gradually return to their alignment with $B_0$, but each proton returns at a different rate. During this period, the gradient coils ($G_x$, $G_y$, and $G_z$) are used to encode spatial information in (d). Finally, the protons return to their alignment with $B_0$ and signal is recorded by receiver coils during that gradual decay in (e).

\begin{table}[t]
\centering
\begin{tabular}{| p{0.15\linewidth} | p{0.15\linewidth} | p{0.38\linewidth} | p{0.18\linewidth}|}

\toprule
\textbf{Component} &
\textbf{Produces} &
\textbf{Primary Function} &
\textbf{Active During} \\
\midrule

Main magnet ($B_0$) &
Strong static magnetic field &
Aligns hydrogen spins and establishes the Larmor frequency &
Entire scan \\ \hline

Transmit RF coil ($B_1$) &
Oscillating radio-frequency magnetic field &
Excites spins by tipping the net magnetization into the transverse plane &
RF excitation \\ \hline

Gradient coils ($G_x,G_y,G_z$) &
Spatially varying magnetic fields &
Encode spatial location by varying the local precession frequency and accumulated phase &
Spatial encoding \\ \hline

Receiver coil &
Induced electrical voltage &
Detects the MR signal emitted by the precessing transverse magnetization &
Signal acquisition \\
\bottomrule
\end{tabular}
\caption{Roles of the major magnetic field components and coils in an MRI system.}
\label{tab:mri_coils}
\end{table}

Suppose the object has transverse magnetization
$m(\mathbf{r})$,
where $\mathbf{r}=(x,y,z)$ denotes the spatial coordinate location.
After the RF excitation pulse, the transverse magnetization continues to precess while gradually decaying. During this period, the receiver coil continuously records the induced electrical voltage as a function of acquisition time $t$, where $t$ denotes the elapsed time after the start of signal acquisition. Because the applied gradient fields vary during signal acquisition, the received MR signal also changes with time and can be expressed as:
\begin{equation}
s(t)
=
\int
m(\mathbf r)
e^{-j2\pi\mathbf k(t)\cdot\mathbf r}
\,d\mathbf r,
\label{eq:mri_signal}
\end{equation}
where the exponential term represents the position-dependent phase accumulated by the spins under the applied gradient fields~\cite{brown2014magnetic}. The vector
\begin{equation}
\mathbf{k}(t)
=
\frac{\gamma}{2\pi}
\int_0^t
\mathbf G(\tau)\,d\tau,
\label{eq:kspace}
\end{equation}
specifies the spatial frequency being sampled at time $t$ and is determined by the applied gradient waveform
$
\mathbf G(t)
=
[G_x(t),G_y(t),G_z(t)]^T
$.
Equation~(\ref{eq:mri_signal}) reveals one of the fundamental properties of MRI. Because the applied gradient fields make the spin phase vary linearly with spatial position, the received MR signal becomes the Fourier transform of the transverse magnetization. Consequently, each measurement recorded by the receiver coil corresponds to a single spatial-frequency sample of the object rather than the signal from a single spatial location~\cite{FesslerMRI}.

\begin{figure}[t]
    \centering
    \includegraphics[width=\linewidth]{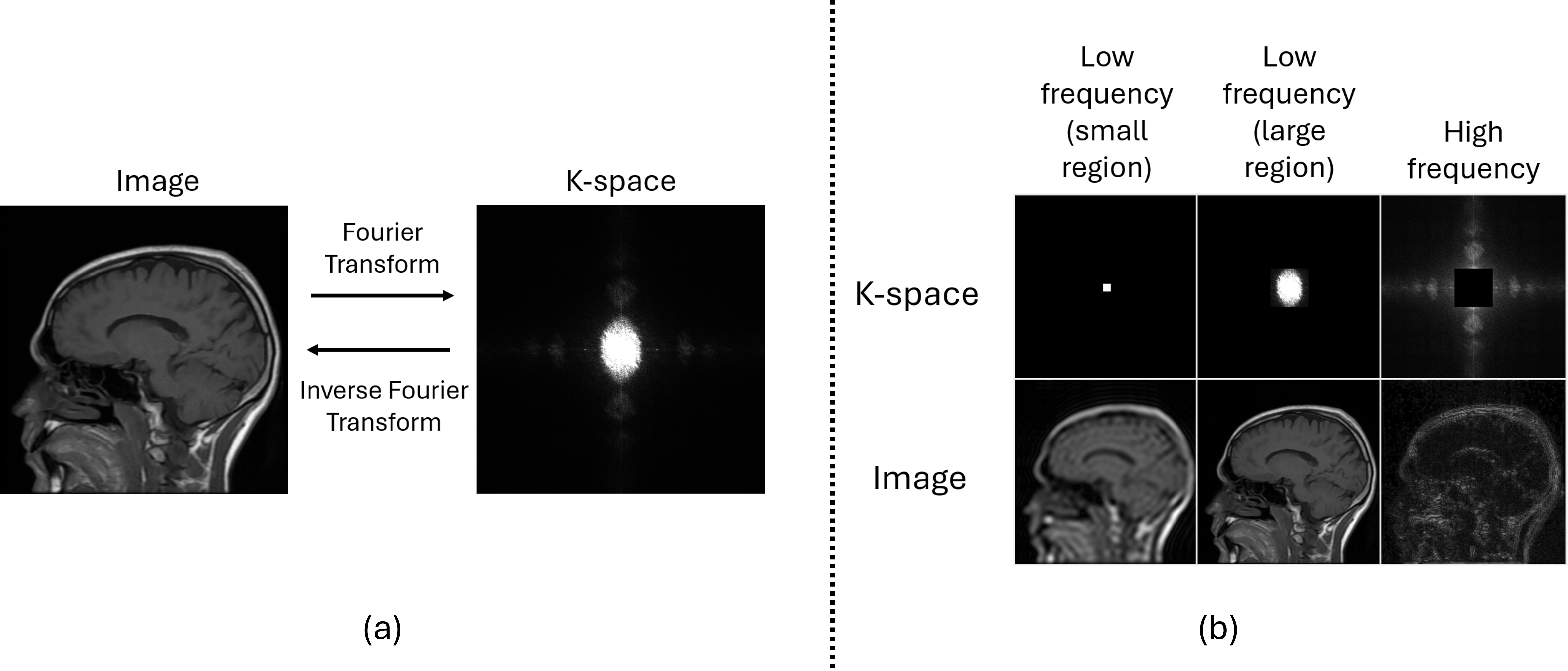}
    \caption{(a) An image of a brain \cite{iflaq2024mri} (left) and its Fourier transform (right). (b) Images of a brain \cite{iflaq2024mri} using a small low frequency region (left), a wider low frequency region (middle), and a high frequency region (right). High frequency region contrast enhanced to improve visualization}
    \label{fig:Fourier}
\end{figure}

During signal acquisition, the scanner continuously varies the gradient waveforms so that measurements are acquired at many different spatial frequencies. The collection of all acquired spatial-frequency measurements is called \emph{k-space}, which is denoted by $\mathbf{Y}$ throughout this chapter.

Unlike the reconstructed image, whose coordinates represent physical locations inside the patient, the coordinates of k-space represent spatial frequencies. For a single receiver coil, if $K_x$, $K_y$, and $K_z$ denote the numbers of sampled spatial frequencies along the three spatial-frequency directions, then the acquired k-space has dimensions $K_x \times K_y$ for two-dimensional imaging and $K_x \times K_y \times K_z$ for three-dimensional imaging.

The k-space representation provides useful physical intuition for MRI
reconstruction. The center of k-space contains low spatial frequencies,
which primarily determines image contrast and large-scale anatomical
structures, whereas the outer regions contain high spatial frequencies
that encode edges and fine structural details. Figure
\ref{fig:Fourier} (a) illustrates an image together with its Fourier
representation, while Figure~\ref{fig:Fourier} (b) demonstrates the effect
of reconstructing images using only the central region, progressively
larger regions, and only the outer regions of k-space \cite{PlewesMRIPrimer}.

\begin{figure}[t]
    \centering
    \includegraphics[width=.9\linewidth]{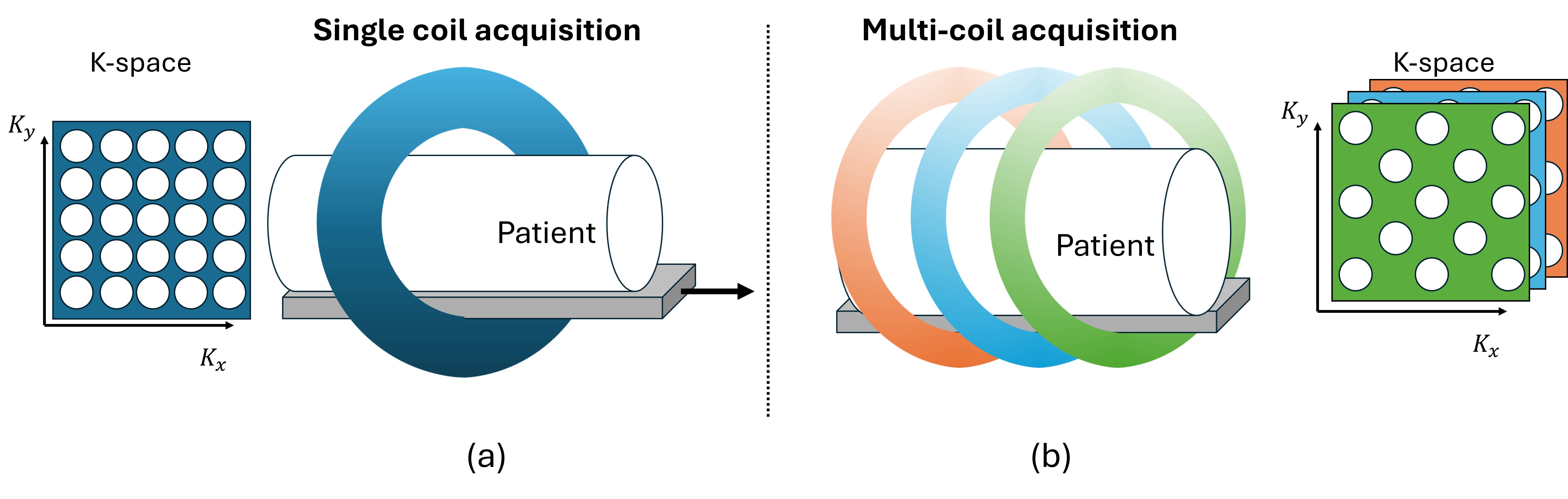}
    \caption{MRI acquisition for single (a) and multi-coil (b) geometries. For single coil acquisitions, k-space must be fully sampled line-by-line leading to longer scans. For multi-coil acquisitions overlapping coils cover the same region and provide additional spatial information that enables under-sampling k-space and thus faster scans without reducing image quality.}
    \label{fig:MR-acquisition modes}
\end{figure}

Modern MRI systems rarely use a single receiver coil. Although a single coil can reconstruct an image, it must acquire a sufficiently dense sampling of k-space to uniquely encode the spatial information of the object, resulting in relatively long scan times. Instead, modern MRI systems employ \emph{parallel imaging}, in which an array of receiver coils simultaneously acquires MR signals~\cite{deshmane2012parallel}.
Each receiver coil has a different spatial sensitivity profile and is most sensitive to signals originating from nearby anatomy~\cite{brown2014magnetic, deshmane2012parallel}. Consequently, the same tissue produces different signal intensities in different receiver coils. These complementary sensitivity profiles provide additional spatial information beyond that encoded by the gradient fields alone. Because part of the spatial encoding is now provided by the receiver coils, the scanner can acquire fewer k-space samples while still reconstructing a high-quality image. As a result, parallel imaging substantially reduces scan time and has become the standard acquisition strategy in modern clinical MRI systems.
Each receiver coil independently acquires its own k-space measurements, introducing an additional coil dimension into the measurement dataset. Figure~\ref{fig:MR-acquisition modes} compares conventional single-coil acquisition (a) with parallel imaging (b) where a single coil covering a full region requires a line-by-line sampling of k-space. Multiple coils for parallel imaging are able to provide additional spatial information which allows the system to under-sample k-space and perform faster scans. Consequently, the acquired data have dimensions
$K_x\times K_y\times N_c$
for two-dimensional imaging and
$K_x\times K_y\times K_z\times N_c$
for volumetric imaging, where
$N_c$
denotes the number of receiver coils.

\begin{table}[htbp]
\centering
\caption{Conceptual comparison of CT, SPECT, and PET data acquisition and their corresponding inverse problems. Although all three modalities measure photons, they reconstruct different physical quantities from different known measurements.}
\label{tab:ct_spect_pet_conceptual}
\renewcommand{\arraystretch}{1.18}
\small
\begin{tabularx}{\textwidth}{
|>{\raggedright\arraybackslash}p{3.2cm}
|>{\raggedright\arraybackslash}X
|>{\raggedright\arraybackslash}X
|>{\raggedright\arraybackslash}X|}
\hline

 & \textbf{CT} & \textbf{SPECT} & \textbf{PET} \\
\hline

\textbf{Imaging principle} &
Measures X-ray attenuation &
Detects single gamma photons emitted by a radiotracer &
Detects coincident $511~\mathrm{keV}$ annihilation photon pairs \\
\hline

\textbf{Signal source} &
External X-ray tube &
Internal gamma-emitting radiotracer &
Internal positron-emitting radiotracer \\
\hline

\textbf{Measured data} &
Transmitted X-ray photon counts &
Detected gamma photon counts &
Detected coincidence photon counts \\
\hline

\textbf{Known during reconstruction} &
Incident X-ray photon counts &
Attenuation map (typically from companion CT) &
Attenuation map (typically from companion CT) \\
\hline

\textbf{Unknown (reconstruction goal)} &
Attenuation coefficient distribution,
$\mu(\mathbf r)$ &
Radiotracer activity distribution,
$\lambda_T(\mathbf r)$ &
Radiotracer activity distribution,
$\lambda_T(\mathbf r)$ \\
\hline

\textbf{Acquisition geometry} &
Rotating X-ray source and detector &
Rotating gamma camera with collimator &
Stationary detector ring with coincidence detection \\
\hline

\textbf{Primary information} &
Structural anatomy &
Functional and molecular activity &
Functional and molecular activity \\
\hline

\textbf{Major strength} &
High spatial resolution and rapid acquisition &
Widely available and relatively inexpensive &
Higher sensitivity and better quantitative accuracy \\
\hline

\textbf{Major limitation} &
Limited functional information &
Low sensitivity due to collimator &
Higher cost and greater system complexity \\
\hline

\textbf{Measurement characteristics} &
High-count, relatively low-noise projections &
Low-count, photon-limited projections &
Higher counts than SPECT but still photon-limited \\
\hline

\textbf{Example sinogram} &
\includegraphics[width=0.95\linewidth,height=0.95\linewidth]{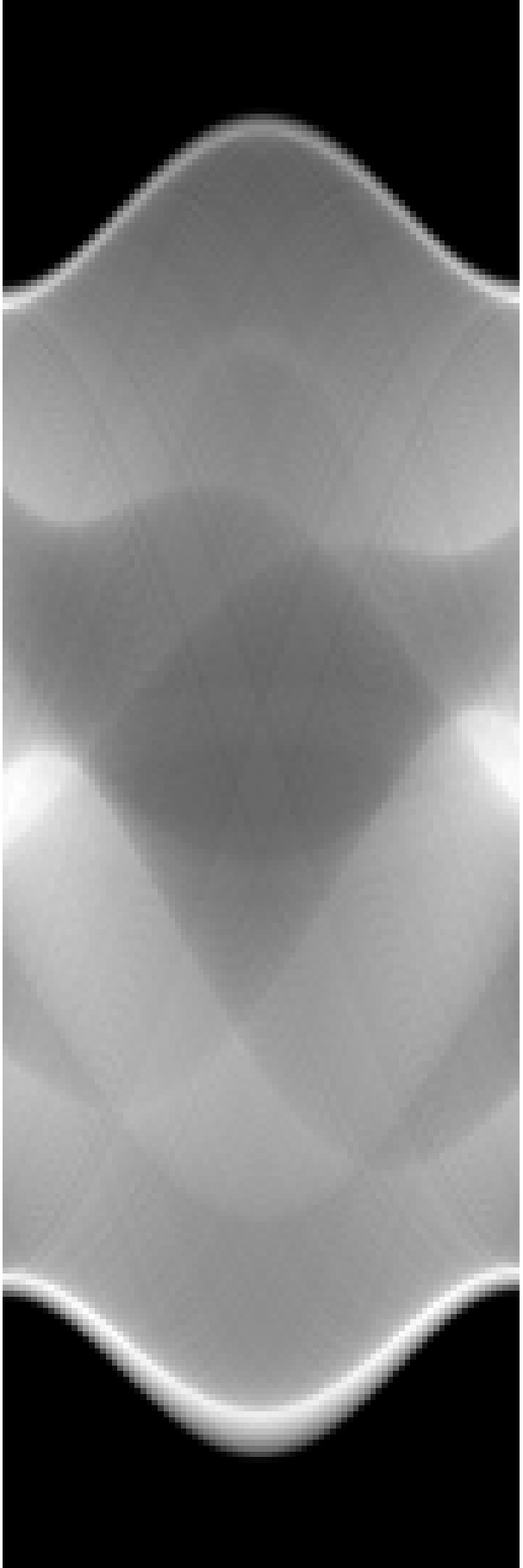}
{\footnotesize Adapted from~\cite{shepp_logan_radon_wikimedia}}&
\includegraphics[width=0.95\linewidth,height=0.95\linewidth,keepaspectratio,angle=90,origin=c]{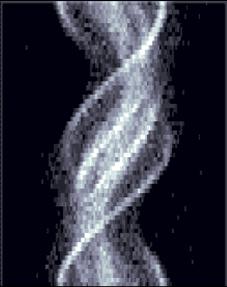}
{\footnotesize Adapted from ~\cite{spect_sinogram_360_wikimedia}.}&
\includegraphics[width=0.95\linewidth,height=0.95\linewidth,keepaspectratio,angle=90,origin=c]{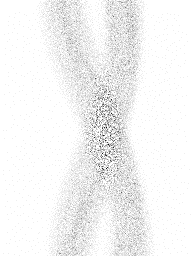}
{\footnotesize Adapted from~\cite{oikonen_pet_reconstruction}}\\
\hline

\textbf{Reconstructed image} &
\includegraphics[width=0.95\linewidth,height=0.95\linewidth]{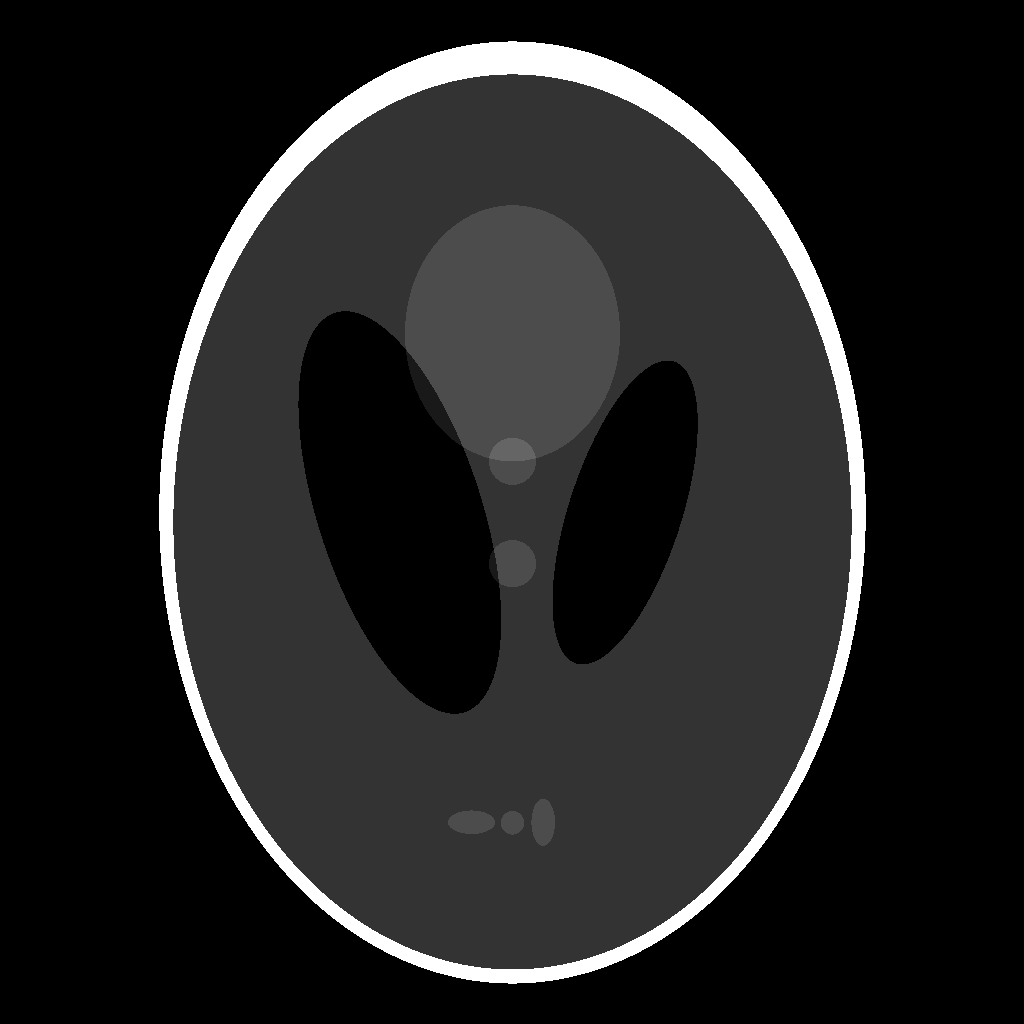}
{\footnotesize Adapted from ~\cite{shepplogan_wikimedia}.} &
\includegraphics[width=0.95\linewidth,height=0.95\linewidth]{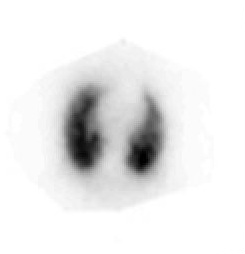}
{\footnotesize Adapted from \cite{spect_lungs_wikimedia}.}&
\includegraphics[width=0.95\linewidth,height=0.95\linewidth]{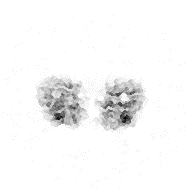}
{\footnotesize Adapted from~\cite{oikonen_pet_reconstruction}}\\
\hline

\end{tabularx}
\end{table}

\subsection{Data Acquisition for Nuclear Medicine Imaging}

CT and MRI primarily visualize anatomical structures, whereas SPECT and PET visualize physiological and molecular processes. After a radiotracer is administered, its uptake reflects biological processes such as metabolism, blood flow, receptor expression, inflammation, oxygen consumption, and bone remodeling. Because these functional changes often occur before structural abnormalities become visible on CT or MRI, SPECT and PET play important roles in early disease detection, staging, treatment planning, and therapy monitoring.

Although CT, SPECT, and PET all measure photons, they solve different inverse problems. In CT, X-ray photons are emitted from an external source, pass through the patient, and are measured after attenuation. Since both the incident X-ray and the transmitted X-ray photon counts are known, the objective is to reconstruct the unknown attenuation coefficient distribution, denoted by $\mu(\mathbf{r})$.

In contrast, SPECT and PET reconstruct the activity distribution, $\lambda_T(\mathbf{r})$, from the injected radiotracer within the patient, and the radiotracer activity distribution determines the expected rate of emitted gamma photons.  As these photons travel through the body, some are attenuated or scattered before reaching the detector, resulting in measured photon counts, denoted by $\lambda_t$. Unlike CT, the attenuation coefficient is not the quantity being reconstructed. Instead, it is typically estimated from a companion CT scan in hybrid SPECT/CT or PET/CT systems and used to correct for photon attenuation. Therefore, the objective of SPECT and PET is to reconstruct the unknown radiotracer activity distribution $\lambda_T(\mathbf r)$ from the detected photon counts using the attenuation map for attenuation correction.

SPECT and PET both reconstruct radiotracer activity distribution but differ fundamentally in how emitted photons are detected. SPECT detects single gamma photons using a mechanical collimator, whereas PET detects pairs of $511~\mathrm{keV}$ annihilation photons in electronic coincidence. Consequently, PET generally provides higher sensitivity and more accurate quantitative imaging, whereas SPECT is less expensive and remains widely available clinically. Table~\ref{tab:ct_spect_pet_conceptual} summarizes the major differences between CT, SPECT, and PET data acquisition.

\subsubsection{Data acquisition for SPECT}

\begin{figure}[htbp]
\centering
\includegraphics[width=\textwidth]{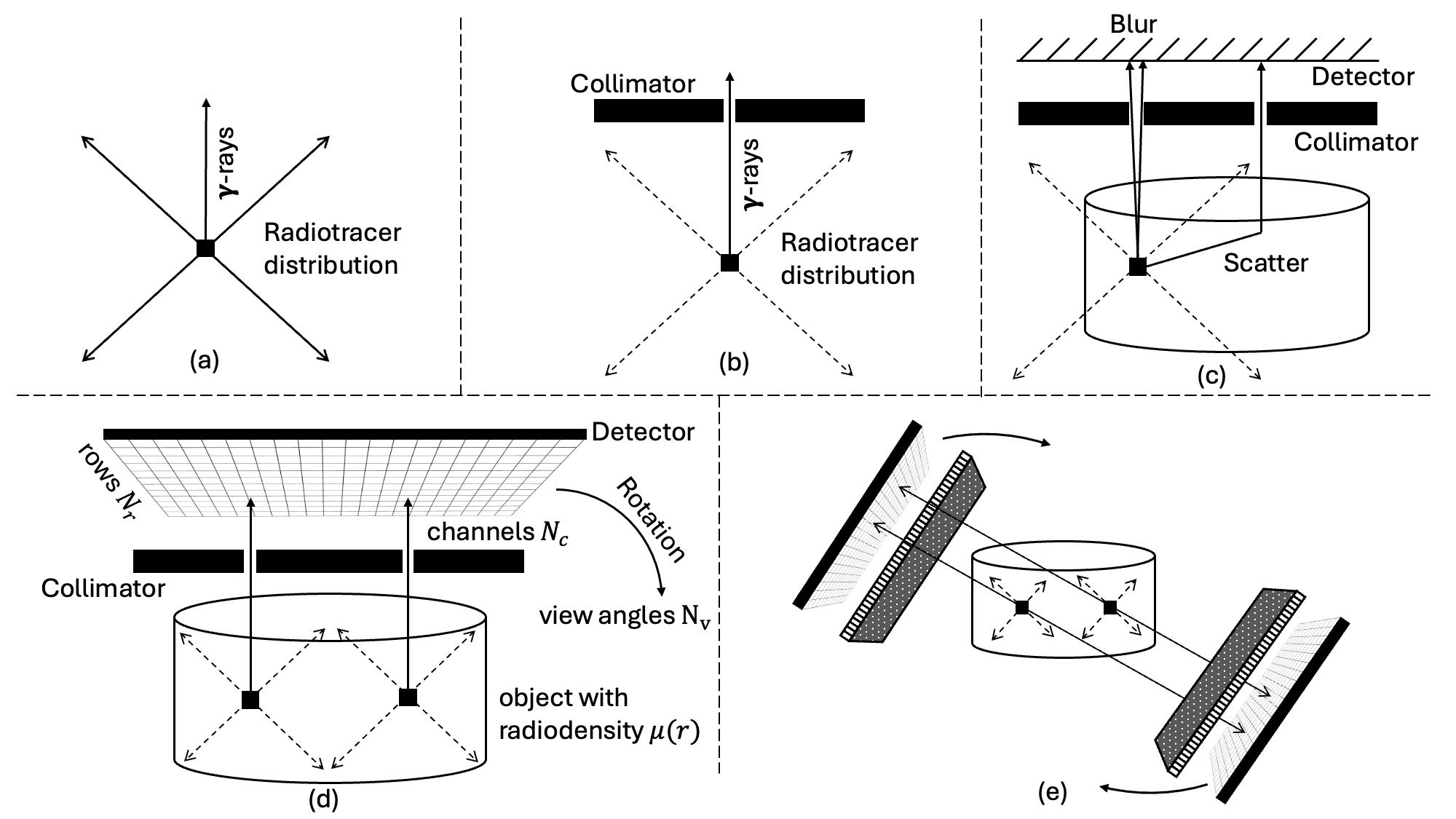}
\caption{(a) Radiotracer activity emits gamma photons omnidirectionally from inside the object, with higher emission in regions of higher uptake. (b) The collimator accepts nearly parallel photons and rejects off-angle photons. (c) Collimator blur occurs because the collimator holes have finite width, allowing photons with slightly different angles to pass through the same hole; scattered photons can also be detected and assigned to incorrect positions. (d) Accepted photons are detected by the detector, where detector rows are represented by $N_r$ and detector columns by $N_c$. (e) Illustration of dual-head SPECT, where two detector heads acquire photons from different angles to increase photon counts and improve acquisition efficiency.}
\label{fig:spect_acquisition}
\end{figure}

The physical principle underlying SPECT is the emission of gamma photons from an injected radiotracer. After administration, the radiotracer accumulates in different tissues according to their physiological function, forming a spatial activity distribution, denoted by $\lambda_T(\mathbf r)$, where the activity represents the expected photon emission rate at each location. Regions with higher radiotracer uptake therefore emit more gamma photons in all directions than regions with lower uptake, as illustrated in Figure~\ref{fig:spect_acquisition}(a).

As the emitted gamma photons travel through the body, some are absorbed or scattered before reaching the detector. This attenuation is described by the linear attenuation coefficient $\mu(\mathbf r)$. Because emitted gamma photons travel in all directions, their direction of origin cannot be inferred directly from the detector measurement. To determine direction, a parallel-hole collimator is placed in front of the detector and it converts photon direction into spatial information by allowing only photons traveling approximately parallel to the collimator holes to reach the detector,  while rejecting most off-angle photons, as shown in Figure~\ref{fig:spect_acquisition}(b)~\cite{Cherry2012PhysicsNuclearMedicine, van2015review}. Because the collimator holes have finite diameter, photons emitted from a single location may be detected by neighboring detector bins, introducing geometric blur shown in Figure~\ref{fig:spect_acquisition}(c). This effect is modeled by the collimator-detector response $C_i(\mathbf r)$. Although the collimator provides the directional information required for tomographic reconstruction, it rejects the vast majority of emitted photons, making SPECT acquisition inherently photon limited and therefore substantially noisier than CT. Combining radiotracer emission, photon attenuation, and the collimator-detector response, the expected detected photon count measured by detector bin $i$ can be written as:

\begin{equation}
\lambda_{t,i}
=
\int_{\Omega}
C_i(\mathbf r)\,
\lambda_T(\mathbf r)\,
e^{
-
\int_{L(\mathbf r,i)}
\mu(\mathbf r')\,d\mathbf r'
}
\,d\mathbf r
+
\epsilon_i ,
\label{eq:spect_single_measurement}
\end{equation}
where $\lambda_{t,i}$ denotes the expected detected photon count measured by detector bin $i$, $C_i(\mathbf r)$ models the combined collimator-detector response, including geometric blur and detector sensitivity, $\lambda_T(\mathbf r)$ is the radiotracer activity distribution, and the exponential term models photon attenuation according to the Beer--Lambert law along the path from $\mathbf r$ to detector bin $i$. The term $\epsilon_i$ represents measurement uncertainty arising primarily from photon-counting statistics and detector noise but also including the photon scattering effect. This is illustrated for detector bin $i$ in Figure~\ref{fig:spect_acquisition}(d).

In addition, it is important to note that SPECT scans do not directly measure the gamma-ray photon attenuation. Instead, the gamma photon attenuation is estimated from a companion CT scan, which provides the X-ray photon attenuation~\cite{fleming1989technique}. The CT-derived attenuation coefficients are then converted to the gamma-ray attenuation  and the estimated gamma photon attenuation is used for the above equation. Consequently, the SPECT attenuation coefficient is treated as known during SPECT reconstruction, while the radiotracer activity remains the unknown image.

During acquisition, one or more gamma-camera detector heads rotate around the patient just like CT. Since the detector rotates around the patient, the detector sensitivity function $C_i(\mathbf r)$ also changes with detector position and acquisition angle.
In addition, the detector is an array of sensors, consisting of
\(N_c\) detector channels in the transverse direction and \(N_r\) detector rows
along the approximate head-to-foot direction. Therefore, each view angle forms a
two-dimensional projection of size \(N_c \times N_r\) illustrated in Figure~\ref{fig:spect_acquisition}(d). Acquiring projections at $N_v$ view angles produces multiple measurements of the same radiotracer distribution from different directions.  Compared with a single-head system, dual-head and multi-head
systems collect more projection data per rotation, increasing photon detection efficiency and
reducing acquisition time, as shown in Figure~\ref{fig:spect_acquisition}(e). 
Each detector view therefore produces a two-dimensional projection containing photon counts measured by every detector element. These projections are analogous to the projection measurements acquired in CT, except that they record detected gamma rather than transmitted X-ray photons.

After projections are acquired from all detector view angles, they are
stacked along the view-angle direction to form the complete SPECT measurement
dataset, denoted as $\mathbf{Y}$. This dataset is the sinogram and has
dimension \(N_c \times N_r \times N_v\). For a
fixed detector row, a two-dimensional slice of \(Y\) is arranged by detector
channels and view angles, where localized radiotracer uptake appears as
sinusoidal traces in the sinogram as the detector rotates around the patient in the same way as the sinusoidal traces for CT. In dynamic SPECT
studies, projections may also be acquired at multiple time points after
radiotracer injection, adding a temporal dimension \(N_t\).

\subsubsection{Data acquisition for PET}

\begin{figure}[!t]
    \centering
    \includegraphics[width=\textwidth]{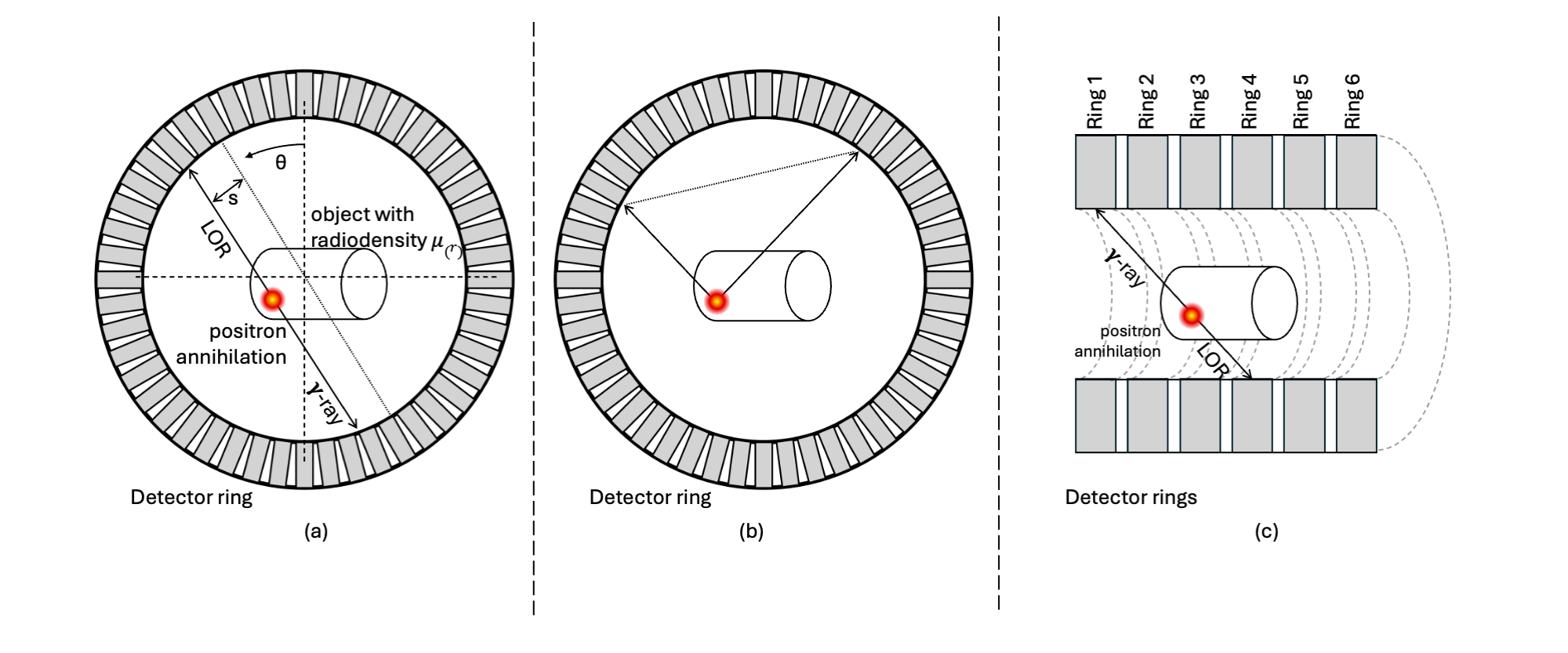}
    \caption{
    (a) A stationary detector ring records annihilation photon pairs and assigns each accepted event to a line of response (LOR), described by angle $\theta$ and radial offset $s$.
    (b) A scattered photon causes the event to be assigned to an incorrect LOR.
    (c) In multi-ring PET, cross-ring coincidences form axial and oblique LORs, providing three-dimensional PET information.}
    \label{fig:pet_geometry}
\end{figure}

The physical principle underlying PET is the detection of pairs of annihilation photons emitted by a positron-emitting radiotracer. After administration, the radiotracer accumulates in different tissues according to their physiological function, forming a spatial activity distribution, denoted by $\lambda_T(\mathbf r)$, where the activity represents the expected positron emission rate at each location. During radioactive decay, each emitted positron travels a short distance before annihilating with an electron, producing two nearly opposite $511~\mathrm{keV}$ photons, as illustrated in Figure~\ref{fig:pet_geometry}(a).

Unlike SPECT, which determines photon direction using a mechanical collimator, PET determines the photon trajectory electronically through \emph{coincidence detection}. If the two annihilation photons are detected by nearly opposite detector elements within a predefined coincidence timing window, the event is accepted as a \emph{true coincidence}, as shown in Figure~\ref{fig:pet_geometry}(a). The two detector elements define a \emph{line of response (LOR)}, denoted by $\mathrm{LOR}_i$, along which the annihilation event is assumed to have occurred. As the photons travel through the patient, they are attenuated according to the linear attenuation coefficient $\mu(\mathbf r)$. In addition, scattered photons, random coincidences, detector normalization, and detector resolution all influence the measured coincidence counts. Combining these effects, the expected coincidence count measured by detector pair $i$ can be expressed as

\begin{equation}
\lambda_{t,i}
=
\int_{\Omega}
C_i(\mathbf r)\,
\lambda_T(\mathbf r)\,
e^{
-
\int_{L(\mathbf r,i)}
\mu(\mathbf r')\,d\mathbf r'
}
\,d\mathbf r
+
\epsilon_i ,
\label{eq:pet_single_measurement}
\end{equation}
where $\lambda_{t,i}$ denotes the expected coincidence count measured by detector pair $i$; $C_i(\mathbf r)$ models the detector-pair sensitivity and normalization; $\lambda_T(\mathbf r)$ is the radiotracer activity distribution to be reconstructed; the exponential term models photon attenuation along the corresponding line of response according to the Beer--Lambert law; and $\epsilon_i$ summarizes measurement uncertainty arising from photon-counting statistics, scattered coincidences, random coincidences, and detector noise.

Compared with SPECT, PET does not require a mechanical collimator or mechanical rotation because coincidence detection directly determines the line of response and stationary detector rings provide angular coverage from all directions. Consequently, PET detects a much larger fraction of emitted photons, resulting in higher system sensitivity, improved quantitative accuracy, and lower image noise. Each accepted coincidence event is assigned to its corresponding line of response \(\mathrm{LOR}_i\).

 After acquisition, all accepted coincidence events are organized into the PET measurement dataset, denoted by $\mathbf Y$. In the conventional \emph{sinogram} representation, coincidence events are grouped according to their corresponding lines of response. For two-dimensional PET imaging, each line of response can be parameterized by a radial offset $s$ and a projection angle $\theta$. Accordingly, each PET projection is stored as a matrix of size $N_c \times N_v$, where $N_c$ denotes the number of radial-offset bins and $N_v$ denotes the number of view-angle bins. 

Modern clinical PET systems employ multiple detector rings along the axial direction. Coincidences detected between different detector rings produce oblique lines of response, enabling fully three-dimensional acquisition, as illustrated in Figure~\ref{fig:pet_geometry}(c). In three-dimensional PET, an additional axial detector-row (ring) dimension is included, giving \(N_c \times N_v \times N_r \times N_{\Delta r}\). Here, \(N_r\) denotes the number of axial detector rings, and \(N_{\Delta r}\) denotes the number of ring-difference bins, where the ring difference is the difference between the two detector-ring indices forming a coincidence line of response. ~\cite{Fahey2002PETDataAcquisition,Defrise2005PETReconstruction}.



In addition to sinogram storage, modern PET systems frequently employ \emph{list-mode acquisition}. Rather than binning coincidence events immediately into a sinogram, list-mode acquisition stores every accepted event individually together with its detector pair, acquisition time, and optionally time-of-flight (TOF) and energy information. Although list-mode data require substantially greater storage and computational resources, they preserve the complete event information and provide greater flexibility for dynamic imaging, cardiac and respiratory gating, motion correction, and time-of-flight PET reconstruction~\cite{Rahmim2005DynamicPET,Zhang2017TotalBodyPET}.

\subsection{Efficient Computing Considerations for Data Acquisition}
Although CT, MRI, PET, and SPECT rely on different imaging physics, the computational challenges of data acquisition share many common characteristics. Modern imaging systems continuously generate large streams of raw detector measurements that must be transferred, buffered, processed, and stored while the scan is in progress. Consequently, data acquisition is not solely a sensing problem but also a real-time computing problem. The acquisition pipeline extends beyond the imaging hardware itself to include detector electronics, communication interfaces, memory systems, storage devices, and data management software. Any bottleneck along this pipeline can reduce acquisition throughput, increase scan time, or degrade image quality through excessive latency or data loss.

Unlike image reconstruction, whose computational cost is dominated by solving large inverse problems, the computational cost of data acquisition is dominated by efficient movement of measurement data. Detector measurements must be streamed continuously from the acquisition hardware into memory and storage with sufficiently high bandwidth and low latency to match the detector output rate. Consequently, efficient computing for data acquisition focuses on maximizing data throughput, minimizing data movement, and reducing unnecessary memory copies rather than maximizing floating-point performance.

These computational considerations strongly influence the design of modern imaging hardware. Rather than relying solely on faster processors, many imaging systems improve acquisition efficiency by increasing hardware parallelism so that more measurements can be acquired simultaneously. In this sense, hardware architecture and efficient computing are tightly coupled. Detector geometry, receiver design, communication interfaces, and memory organization are all optimized to maximize acquisition throughput while minimizing scan time and data movement.

Computed tomography provides a representative example. Early CT scanners acquired one detector row at a time, requiring multiple gantry rotations to image a large anatomical volume. Modern multidetector CT systems instead employ tens to hundreds of detector rows, allowing many slices to be acquired simultaneously during each rotation and substantially increasing acquisition efficiency~\cite{kalender2011computed}. More recently, photon-counting CT detectors record not only the arrival of each X-ray photon but also its energy, enabling spectral CT imaging and improved material discrimination~\cite{taguchi2017energy}. Although this additional spectral information significantly increases data rates, it also motivates efficient front-end hardware, FPGA-based processing, high-bandwidth communication, and memory-efficient data streaming to sustain real-time acquisition.

MRI employs a different strategy for improving acquisition efficiency. As discussed in Section~\ref{sec:MRIdata}, modern MRI systems rarely rely on a single receiver coil. Instead, they use parallel imaging, where arrays of receiver coils simultaneously acquire MR signals from different spatial locations~\cite{SENSE,GRAPPA}. Because each receiver coil provides complementary spatial information through its sensitivity profile, fewer k-space measurements are required to reconstruct the image, thereby reducing scan time. However, parallel imaging also produces multiple synchronized streams of complex-valued k-space data that must be transferred, buffered, and synchronized efficiently. Consequently, efficient memory management, multi-channel communication, and low-latency data streaming become essential components of modern MRI acquisition systems.

Nuclear medicine imaging follows a similar philosophy. In SPECT, modern systems often employ dual-head or triple-head gamma cameras so that multiple projection views are acquired simultaneously, reducing acquisition time and improving photon detection efficiency~\cite{Cherry2012PhysicsNuclearMedicine}. PET scanners achieve similar improvements by surrounding the patient with multiple rings of detector modules that detect coincidence events from many directions simultaneously. Modern digital PET systems further increase detector count rates and timing resolution, generating enormous streams of photon detection events that must be processed, coincidence matched, and transferred in real time. Consequently, efficient communication, parallel event processing, and hardware-aware data management are critical for sustaining modern PET acquisition rates.

Although the acquisition mechanisms differ across imaging modalities, they share a common computational objective: maximize the amount of useful measurement information acquired per unit time while minimizing unnecessary data movement, latency, and hardware cost. Efficient computing therefore begins at the hardware and system-design level, where detector geometry, receiver architecture, memory organization, and communication bandwidth are co-designed with acquisition algorithms. These design choices ultimately determine how efficiently measurement data can be delivered to the reconstruction pipeline. In the following sections, the computational bottleneck shifts from efficient data movement during acquisition to efficient numerical computation for image reconstruction.

\section{Image Reconstruction}
\label{sec:image reconstruction}

\subsection{General Framework}
\label{subsec:general_framework}

The objective of medical image reconstruction is to recover a clinically interpretable image from the raw measurements acquired during data acquisition. As explained in Section~\ref{sec:data acquisition}, CT acquires X-ray projections, MRI samples k-space, and PET and SPECT detect emitted photons from radioactive tracers. Although these imaging modalities rely on different physical principles, they all share the same computational objective: estimating an unknown image from indirect measurements. Consequently, image reconstruction across a wide range of imaging modalities can be described within a common mathematical framework.

The relationship between the acquired measurements and the unknown image is commonly described by the following \emph{forward imaging model}:

\begin{equation}
\mathbf{Y}
=
\mathbf{A}\mathbf{X}
+
\mathbf{E},
\label{eq:y=ax}
\end{equation}
where the forward system matrix operator $\mathbf{A}$ models how the imaging system transforms the unknown image into measurable data, and $\mathbf{E}$ represents measurement uncertainty, including electronic noise, photon-counting statistics, detector imperfections, and other sources of acquisition error. Image reconstruction seeks to estimate the unknown image $\mathbf{X}$ from the acquired measurements $\mathbf{Y}$ by approximately inverting this forward imaging model.

Equation~(\ref{eq:y=ax}) provides a unified mathematical framework for nearly all medical image reconstruction algorithms discussed in this chapter. Although CT, MRI, PET, and SPECT rely on different imaging physics, they can all be interpreted as different implementations of the same forward imaging model. The primary differences among reconstruction algorithms lie in how they model the forward operator $\mathbf{A}$, how they characterize the measurement uncertainty $\mathbf{E}$, and how they solve the resulting inverse problem.

The unknown image volume is denoted by
$\mathbf{X}\in\mathbb{R}^{N}$,
where $N$ is the total number of voxels in the reconstructed image. Although the mathematical notation is identical across imaging modalities, the physical meaning of $\mathbf{X}$ depends on the imaging physics. For CT, $\mathbf{X}$ represents the linear attenuation coefficient distribution, $\mu(\mathbf{r})$. For MRI, it represents the complex-valued transverse magnetization distribution, $m(\mathbf{r})$. For PET and SPECT, it represents the radiotracer activity distribution, $\lambda_T(\mathbf{r})$.

The measurement vector is denoted by
$\mathbf{Y}\in\mathbb{R}^{M}$,
where $M$ is the total number of acquired measurements. Throughout this section, the measurements are vectorized for notational convenience, although they are naturally stored as multidimensional arrays such as sinograms or k-space. For CT, PET, and SPECT, $\mathbf{Y}$ corresponds to the acquired sinogram, with
$
M=N_v\times N_r\times N_c,
$
where $N_v$ is the number of projection views and $N_r\times N_c$ is the detector size. For MRI, $\mathbf{Y}$ represents the acquired k-space measurements, with
$
M=K_x\times K_y\times K_z\times N_c,
$
where $K_x$, $K_y$, and $K_z$ denote the sampled k-space dimensions and $N_c$ is the number of receiver coils.

The forward operator $\mathbf A$ describes how the imaging system transforms an image into measurable data. It is the central mathematical component of image reconstruction because it encapsulates the imaging physics, scanner geometry, detector response, and sampling strategy. The mathematical representation and computational implementation of $\mathbf A$ depend on both the imaging modality and the reconstruction algorithm.
In CT, PET, and SPECT, $\mathbf{A}$ is naturally interpreted as a system matrix, whose elements quantify the contribution of each image voxel to each detector measurement according to the imaging geometry and physical model.  Depending on the imaging geometry and reconstruction algorithm, this matrix may be explicitly stored, represented in compressed form, or evaluated implicitly using projector algorithms.
In MRI, $\mathbf{A}$ is naturally expressed as a composition of linear operators, including coil sensitivity weighting, Fourier encoding, and k-space sampling~\cite{FesslerMRI}. Rather than explicitly constructing a matrix, MRI reconstruction typically evaluates these operators directly using Fast Fourier Transforms (FFTs) together with coil sensitivity maps. Regardless of how $\mathbf{A}$ is represented, image reconstruction seeks to estimate the unknown image $\mathbf{X}$ by approximately inverting the forward model in Equation~\eqref{eq:y=ax}. The choice of how accurately $\mathbf{A}$ models the imaging physics largely determines both the reconstruction quality and the computational complexity.

Image reconstruction is fundamentally an inverse problem because the unknown image $\mathbf{X}$ must be estimated from indirect and imperfect measurements $\mathbf{Y}$. In principle, if $\mathbf A$ were square, perfectly known, and invertible, the image $\mathbf{x}$ could be recovered by directly computing $\mathbf A^{-1}\mathbf Y$. In practice, however, $\mathbf A$ is often ill-conditioned, non-square, or too large to invert explicitly, making direct inversion impractical~\cite{bouman2022foundations}. Unlike well-conditioned linear systems encountered in linear algebra, medical image reconstruction is frequently \emph{ill-posed}, meaning that small perturbations in the measurements may produce large changes in the reconstructed image, and in some situations multiple images may satisfy the measurement equation equally well. Three major challenges contribute to this difficulty.

The first challenge is incomplete data acquisition. In many clinical applications, the measurement vector $\mathbf{Y}$ is intentionally undersampled to reduce radiation dose, shorten scan time, improve patient comfort, reduce motion artifacts, or increase clinical throughput. Examples include sparse-view CT, accelerated fast scan MRI using k-space undersampling, and reduced-count PET and SPECT imaging. From a linear algebra perspective, undersampling reduces the number of measurements while the number of unknown image voxels continues to increase as image resolution improves. Consequently, the system becomes underdetermined, allowing many possible image volumes to satisfy the measurement equation. Additional assumptions, prior knowledge, or regularization are therefore required to recover a unique and physically meaningful solution.

The second challenge is measurement uncertainty. The measurement uncertainty $\mathbf{E}$ cannot be eliminated because it is intrinsically coupled to important clinical considerations. In CT, lowering the radiation dose reduces the number of detected X-ray photons, increasing measurement noise while decreasing the patient's exposure to ionizing radiation. In MRI, shorter scan times produce fewer acquired measurements or lower signal-to-noise ratio but improve patient comfort, reduce motion artifacts, and increase scanner utilization, particularly for pediatric, elderly, and critically ill patients. Similarly, PET and SPECT often seek to reduce radiotracer dose or acquisition time to improve patient safety and clinical efficiency, both of which increase measurement uncertainty. Consequently, reconstruction algorithms must recover diagnostically useful images from noisy and imperfect measurements.

The third challenge is computational complexity. Modern imaging systems routinely reconstruct three-dimensional or four-dimensional image volumes containing millions to billions of voxels, while the acquired measurements may contain hundreds of millions or even billions of samples. Although Equation~(\ref{eq:y=ax}) is written in matrix form, explicitly storing or directly inverting the forward operator is often computationally infeasible for modern imaging systems. Instead, reconstruction algorithms evaluate the action of $\mathbf{A}$ and its adjoint implicitly through computational procedures such as forward projection, back-projection, or Fourier encoding. These operators are repeatedly evaluated throughout reconstruction and therefore dominate the computational cost. Their computational efficiency, memory usage, numerical accuracy, and algorithmic convergence rate largely determine reconstruction speed and image quality.

Different reconstruction algorithms address the inverse problem in Equation~(\ref{eq:y=ax}) by making different assumptions about the forward
operator $\mathbf{A}$ and the statistical properties of the unknown image and measurements. As shown in Table~\ref{tab:reconstruction_comparison},
these methods differ primarily in two aspects: (1) the fidelity of the
forward imaging model, and (2) whether Bayesian inference is employed to incorporate measurement statistics and prior knowledge.

Analytical reconstruction methods, such as Filtered Backprojection (FBP) for CT and inverse Fourier reconstruction for MRI~\cite{feldkamp1984practical,hoffman2016freect_wfbp, aibinu2008mri}, assume simplified forward imaging models for which approximate closed-form inverse solutions can be derived. These methods achieve extremely high
computational efficiency and remains the standard reconstruction approach for many clinical imaging systems.

Iterative reconstruction methods abandon closed-form inversion and
instead formulate image reconstruction as an optimization problem. They
repeatedly evaluate the forward operator $\mathbf{A}$ and its adjoint to progressively refine the image estimate. To reduce computational cost, the $\mathbf{A}$ matrix is often represented using computationally efficient approximations of the imaging physics while maintaining acceptable reconstruction accuracy.

Statistical iterative reconstruction methods, also known as \emph{Model-Based Iterative Reconstruction} (MBIR)~\cite{wang2016high,wang2019consensus,wang2021physics,bouman2022foundations,FesslerMRI} extends iterative reconstruction by incorporating statistical models of the measurements together with prior knowledge of the reconstructed image. These methods typically employ more accurate forward models and Bayesian optimization to improve reconstruction quality under low-dose, sparse-view, and noisy acquisition conditions. At the same time, however, their substantially higher computational cost motivates many of the efficient computing techniques discussed later in this chapter.

The following sections review these reconstruction methods in detail and discuss how advances in efficient computing and HPC enable increasingly sophisticated image reconstruction algorithms while maintaining clinically practical reconstruction times.

\begin{table*}[t]
\centering
\caption{Comparison of major medical image reconstruction methods.}
\label{tab:reconstruction_comparison}
\small
\resizebox{\linewidth}{!}
{
\begin{tabular}{|p{0.17\linewidth} |
                p{0.17\linewidth} |
                p{0.10\linewidth} |
                p{0.10\linewidth} |
                p{0.10\linewidth} |
                p{0.10\linewidth} |
                p{0.09\linewidth} |
                p{0.12\linewidth}|} 
\toprule

\textbf{Categories}
&
\textbf{Forward Operator ($\mathbf{A}$)}
&
\textbf{Bayesian}
&
\textbf{Handles Sparse Data}
&
\textbf{Handles Noise}
&
\textbf{Compute Cost}
&
\textbf{Clinical Use}
&
\textbf{Methods}
\\
\midrule

Analytical
&
Simplified closed-form analytical inverse
&
No
&
Poor
&
Poor
&
Low
&
Very High
&
FBP, FDK, Inverse FFT
\\ \hline

Iterative Reconstruction
&
Simplified approximated projector model
&
No
&
Moderate
&
Moderate
&
Medium
&
High
&
ART, SART
\\ \hline

Statistical Iterative Reconstruction
&
High-fidelity system model
&
Yes
&
Good
&
Good
&
Very High
&
Growing
&
PWLS, MBIR
\\

\bottomrule
\end{tabular}
}
\end{table*}

\subsection{Analytical Reconstruction Methods}

Analytical reconstruction methods recover the unknown image by exploiting closed-form mathematical inverses of the forward imaging model. Because these methods avoid iterative optimization, they are computationally efficient and have long served as the foundation of clinical image reconstruction. Among all medical imaging modalities, CT and MRI possess particularly elegant analytical formulations. In CT, the forward operator is described by the Radon transform, whose inverse is efficiently approximated by Filtered Backprojection (FBP)~\cite{kak1988principles}. In MRI, the forward operator is the Fourier transform, allowing image reconstruction through the inverse Fourier transform~\cite{aibinu2008mri,PlewesMRIPrimer}. Consequently, FBP for CT and inverse Fourier reconstruction for MRI remain the standard analytical reconstruction methods used in clinical practice.

Analytical reconstruction methods have also been developed for PET and SPECT, primarily based on Radon transform formulations and FBP. However, unlike CT, the forward operators $\mathbf{A}$ for PET and SPECT must model additional imaging physics for nuclear medicine imaging and photon scattering. These effects cannot be accurately represented by the simple Radon transform alone. Consequently, analytical reconstruction methods generally produce lower image quality for PET and SPECT, and these modalities transitioned to iterative and statistical iterative reconstruction methods much earlier than CT.

\begin{figure}[h]
\centering
\includegraphics[width=.35\textwidth]{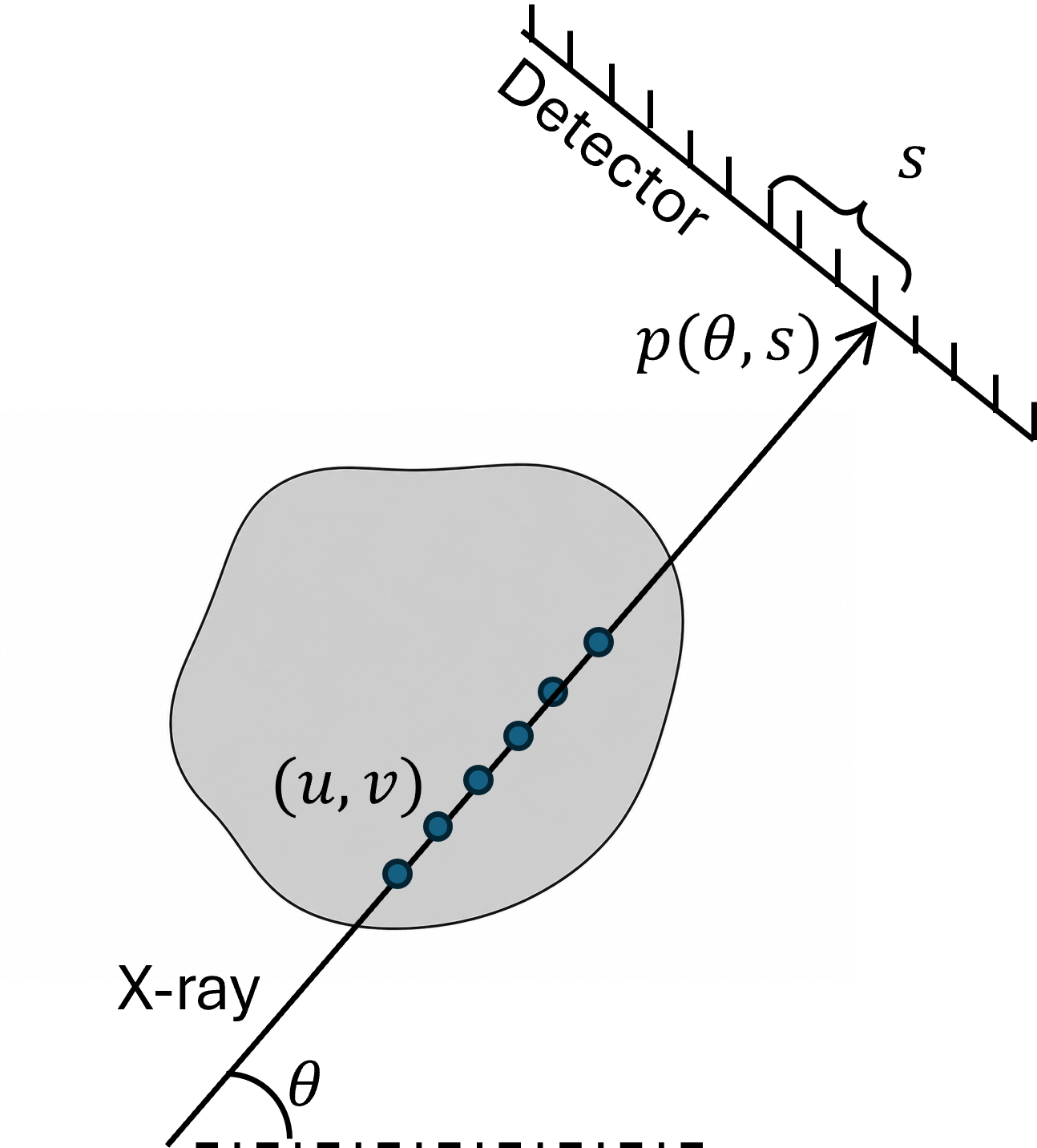}
\caption{Illustration of the Radon transform. For a projection view angle $\theta$, detector channel $s$ measures the line integral of the attenuation coefficient distribution $\mathbf{X}$ along the corresponding X-ray path.}
\label{fig:radon.geometry}
\end{figure}

\subsubsection{Radon Transform Analytical Reconstruction for CT}
\label{subsec:FBP}

For X-ray CT, the forward imaging operator introduced in Equation~(\ref{eq:y=ax}) has a particularly elegant analytical representation known as the \emph{Radon transform}~\cite{kak1988principles}. The Radon transform maps the unknown linear attenuation coefficient distribution $\mathbf{X}$ into the sinogram $\mathbf{Y}$. Consequently, CT reconstruction can be formulated as the inverse problem of recovering $\mathbf{X}$ from its Radon transform.

For simplicity, consider a two-dimensional parallel-beam CT system. Let
$x(u,v)$ denote the linear attenuation coefficient at the spatial coordinate
$(u,v)$ within the object. Note that for analytical reconstruction, $x(u,v)$ is a continuous function for the attenuation coefficient distribution. This corresponds to the vector $\mathbf{X}$ in the general computing framework. At a projection view angle $\theta$, the detector consists of multiple detector channels, as illustrated in Figure~\ref{fig:radon.geometry}. Each detector channel records the X-ray measurement arriving at a different location along the detector.

To describe the detector mathematically, we introduce a continuous detector coordinate $s$, which specifies the location along the detector measured relative to its center. In a practical CT scanner, the detector is discretized into finite detector channels, each corresponding to a small interval of detector coordinates. Thus, the detector coordinate $s$ identifies the detector channel at which the X-ray is measured. For a fixed view angle $\theta$, the X-ray corresponding to detector coordinate $s$ satisfies the following line equation~\cite{kak1988principles}:

\begin{equation}
-u\sin\theta+v\cos\theta=s \ .
\label{eq:line}
\end{equation}

As discussed in Section~\ref{subsec:ct-data-acquisition}, the detector measures the accumulated attenuation along an X-ray path. Therefore, the detector measurement at view angle $\theta$ and detector coordinate $s$ is given by the line integral: 
\begin{equation}
p(\theta,s)
=
\int_{L(\theta,s)}
x(u,v)\, dl,
\end{equation}
where $L(\theta,s)$ denotes the X-ray path corresponding to detector coordinate $s$, and $dl$ represents an infinitesimal distance along the ray. Note that for analytical reconstruction, $p(\theta,s)$ is a continuous function for detector measurement, and it corresponds to the vector $\mathbf{Y}$ in the general computing framework.

The line integral can be rewritten as an equivalent two-dimensional integral using the Dirac delta function. This equivalent formulation is known as the Radon transform~\cite{kak1988principles}:

\begin{equation}
p(\theta,s)
=
\int\!\!\int
x(u,v)\,
\delta\!\left(
s+u\sin\theta-v\cos\theta
\right)
\,du\,dv,
\label{eq:radon}
\end{equation}
where $\delta(\cdot)$ denotes the Dirac delta function. The delta function acts as a selector that restricts the integration to points satisfying Equation~(\ref{eq:line}), namely the X-ray corresponding to detector coordinate $s$. Repeating Equation~\eqref{eq:radon} over all detector coordinates and all view angles produces the complete sinogram. Therefore, the Radon transform provides the analytical expression of the CT forward operator $\mathbf{A}$ introduced in Equation~\eqref{eq:y=ax}. 

With the above equation, the reconstruction is then to inverse the Radon transform. Given the measured sinogram $\mathbf{Y}$, the objective is to recover the attenuation coefficient distribution $\mathbf{X}$ by computing the \emph{inverse Radon transform}. Under ideal mathematical assumptions that the projections are continuous functions of detector position and are noise-free and there are infinitely many view angles, the inverse Radon transform admits an exact closed-form solution. In practice, however, CT measurements are discrete, noisy, and acquired from a finite number of detector channels and view angles. Consequently, directly evaluating the exact inverse Radon transform is neither practical nor robust for clinical CT systems.

Instead, practical CT scanners employ the \emph{Filtered Backprojection (FBP)} algorithm~\cite{Chen2021PortableBP,Stierstorfer_2004} and its extensions, such as the Feldkamp--Davis--Kress (FDK) algorithm for cone-beam CT~\cite{feldkamp1984practical,chen2019ifdk,chen2021scalable}. FBP is an analytical reconstruction method that provides an effective approximation to the inverse Radon transform while remaining computationally simple and highly efficient.

\begin{figure}[h]
\centering
\includegraphics[width=\textwidth]{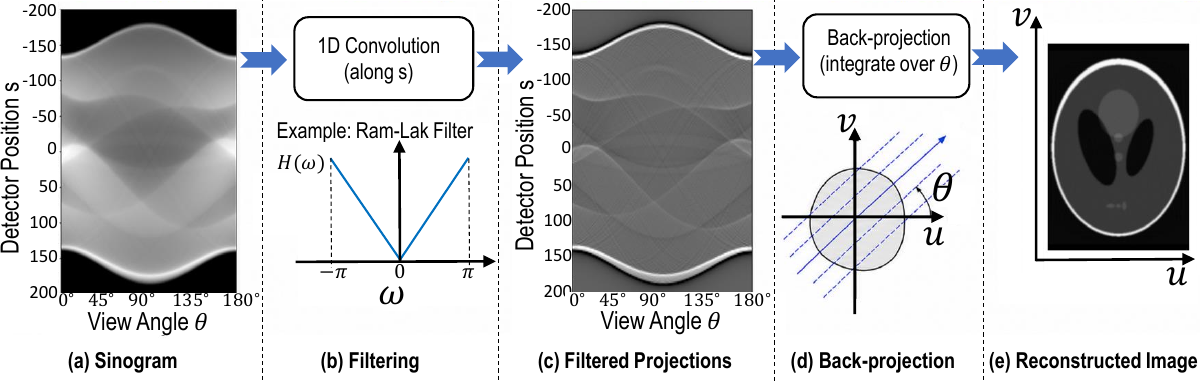}
\caption{FBP reconstruction process. (a) Projection data are acquired over multiple view angles to form a sinogram. (b) Each projection is filtered along the detector direction (namely, $s$) using a reconstruction filter (e.g., Ram--Lak $H(\omega)$~\cite{kak2001principles}) to compensate for back-projection blurring. (c) The filtered projections are generated. (d) The filtered projections are backprojected along the corresponding X-ray paths and accumulated over all view angles (namely, $\theta$). (e) The reconstructed two-dimensional image is obtained.}
\label{fig:fbp.pipeline}
\end{figure}

The basic idea of FBP is illustrated in Figure~\ref{fig:fbp.pipeline}. During CT acquisition, each detector measurement $p(\theta,s)$ is obtained by integrating the attenuation coefficients along one X-ray path through the object. Reconstruction reverses this process. For each measurement, the measured projection value is distributed back along the same X-ray path that produced it. This operation is called \emph{back-projection}. Because the detector measures only the total attenuation along an X-ray path rather than the attenuation of individual voxels, the same projection value is assigned to every voxel lying on that ray. Repeating this process for all detector coordinates and all view angles causes contributions from many X-rays to intersect and accumulate, gradually reconstructing the attenuation coefficient distribution.

However, back-projection alone does not exactly recover the original image. Since each projection value is uniformly spread along an entire X-ray path, neighboring X-rays produce overlapping contributions, resulting in a blurred reconstruction. To compensate for this blur, each projection is first filtered along the detector direction before back-projection. This one-dimensional filtering operation compensates for the low-pass blurring introduced by back-projection by amplifying high spatial frequencies and is commonly implemented using filters such as the Ram--Lak filter. The filtered projections are then backprojected and accumulated over all view angles. Because the reconstruction consists of a filtering step followed by back-projection, the algorithm is known as FBP.

Mathematically, the FBP reconstruction is expressed as~\cite{kak1988principles}:
\begin{equation}
x(u,v)
=
\int_0^\pi
\left[
p(\theta,s)*h(s)
\right]_{s=-u\sin\theta+v\cos\theta}
d\theta,
\label{eq:fbp}
\end{equation}
where $h(s)$ denotes the reconstruction filter and $*$ represents one-dimensional convolution along the detector direction.

From the perspective of the general reconstruction framework in Section~\ref{subsec:general_framework}, FBP replaces the large forward operator $\mathbf{A}$ with the analytical Radon transform and its inverse. Rather than constructing an explicit system matrix or repeatedly solving an optimization problem, FBP exploits the mathematical properties of the Radon transform to derive a closed-form reconstruction algorithm. Consequently, FBP has very low computational complexity, requires relatively little memory, and remains one of the fastest reconstruction algorithms used in clinical CT.

The simplicity of FBP, however, comes at the expense of reconstruction accuracy. Because the Radon transform is an idealized forward model, FBP assumes complete and uniformly sampled projection data together with simplified imaging physics. It does not accurately model detector response, scattering, beam hardening, or other physical effects present in modern CT systems~\cite{wang2021physics}. Consequently, FBP addresses the three reconstruction challenges introduced in Section~\ref{subsec:general_framework} only to a limited extent. It is sensitive to measurement noise in low-dose CT, performs poorly for sparse-view and limited-angle acquisitions, and cannot easily incorporate more accurate imaging models or prior information. Nevertheless, owing to its closed-form solution, high computational efficiency, and robust performance under standard acquisition protocols, FBP remains one of the most widely used reconstruction methods in clinical CT and serves as the foundation upon which modern iterative and statistical reconstruction methods have been developed.

\subsubsection{Fourier Analytical Reconstruction for MRI}
\label{sec:MRI_analytic}

As discussed in Section~\ref{sec:MRIdata},  MRI acquisition directly samples the Fourier transform of the transverse magnetization. During acquisition, each measurement corresponds to one spatial-frequency sample in k-space, and the complete k-space dataset is denoted by $\mathbf{Y}$. Consequently, MRI reconstruction is fundamentally the inverse problem of recovering the image from its sampled Fourier transform.

For an ideal MRI system with complete Cartesian sampling, negligible field inhomogeneity, and a single receiver coil, the reconstruction problem has a simple analytical solution. Since the acquired k-space measurements are the Fourier transform of the transverse magnetization distribution $m(\mathbf{r})$, the image can be reconstructed in closed-form by applying the inverse Fourier transform to the measured k- space~\cite{brown2014magnetic}:

\begin{equation}
m(\mathbf r)
=
\mathcal{F}^{-1}
\{\mathbf Y\} \ ,
\label{eq:IFFT}
\end{equation}
where $\mathcal{F}^{-1}$ denotes the inverse Fourier transform operation.
In practice, the inverse Fourier transform is implemented using the \emph{inverse Fast Fourier Transform} (IFFT), which reduces the computational complexity from
$O(N^2)$
to
$O(N\log N)$.
This dramatic reduction in computational cost makes MRI reconstruction highly efficient and enables near-real-time image reconstruction in modern clinical scanners that have limited computing resources.

Although IFFT reconstruction is computationally simple, it relies on several idealized assumptions. It assumes that k-space has been fully sampled on a Cartesian grid, that the magnetic field is perfectly homogeneous, and that the measurements are acquired from a single receiver coil. These assumptions closely match the physical design of early MRI systems, which is why inverse Fourier reconstruction became the standard clinical reconstruction algorithm. Modern MRI systems, however, generally violate these assumptions. Clinical scanners routinely employ arrays of multiple receiver coils for parallel imaging~\cite{deshmane2012parallel}, non-Cartesian sampling trajectories such as radial or spiral acquisition~\cite{feng2014golden}, and accelerated imaging through k-space undersampling. In addition, magnetic field inhomogeneity, gradient imperfections, and relaxation effects introduce deviations from the ideal Fourier model \cite{FesslerMRI}.

From the perspective of the general reconstruction framework introduced in Section~\ref{subsec:general_framework}, the forward operator $\mathbf{A}$ for an ideal MRI system is simply the Fourier transform, and analytical reconstruction computes its inverse through the inverse Fourier transform. Unlike CT, where FBP provides an approximation to the inverse Radon transform, the inverse Fourier transform is mathematically exact for fully sampled Cartesian acquisitions. Similar to FBP, however, analytical Fourier reconstruction is computationally highly efficient because it exploits the mathematical structure of the imaging model rather than solving an optimization problem.

The principal limitation of analytical Fourier reconstruction is that it relies on the ideal Fourier acquisition model. In practice, modern MRI systems employ multiple receiver coils, accelerated k-space undersampling, non-Cartesian sampling trajectories, and are affected by field inhomogeneity, gradient imperfections, relaxation effects, and measurement noise. These factors cannot be fully addressed by a simple inverse Fourier transform. Consequently, modern MRI reconstruction increasingly employs iterative, statistical, and AI-based reconstruction methods that explicitly model these physical effects to improve image quality, particularly for accelerated and high-resolution imaging.

\subsection{Iterative and Statistical Iterative Reconstruction Methods}

\subsubsection{Iterative Reconstruction Methods for CT}
\label{sec:iterative_ct}

FBP reconstructs CT images by analytically approximating the inverse Radon transform. As discussed in the previous subsection~\ref{subsec:FBP}, this approach derives a closed-form inverse assuming the projection data are represented as continuous functions of detector position and projection view angle. In practice, however, CT scanners acquire only discrete projection measurements at finite detector channels and finite projection view angles. Consequently, FBP reconstructs the image by interpolating these discrete measurements and numerically approximating the continuous inverse Radon transform. When the projection sampling becomes sparse, the interpolation becomes inaccurate, leading to streak artifacts, aliasing, and loss of spatial resolution. Furthermore, because the Radon transform models only an idealized imaging system, FBP cannot easily incorporate more accurate models of detector response, focal spot blur, beam hardening, scattering, or other physical effects that influence image quality.

To overcome these limitations, iterative reconstruction, such as ART~\cite{gordon1974tutorial} and its variants such as SIRT and SART~\cite{andersen1984simultaneous}, abandons the continuous Radon transform formulation and instead models CT acquisition directly in the discrete domain. Rather than approximating the forward operator $\mathbf{A}$ using analytical transforms, iterative reconstruction constructs the forward operator as a discrete system matrix and explicitly describes the relationship between every discrete detector measurement and every image voxel through linear algebra.

\begin{figure}[t]
\centering
\includegraphics[width=.6\textwidth]{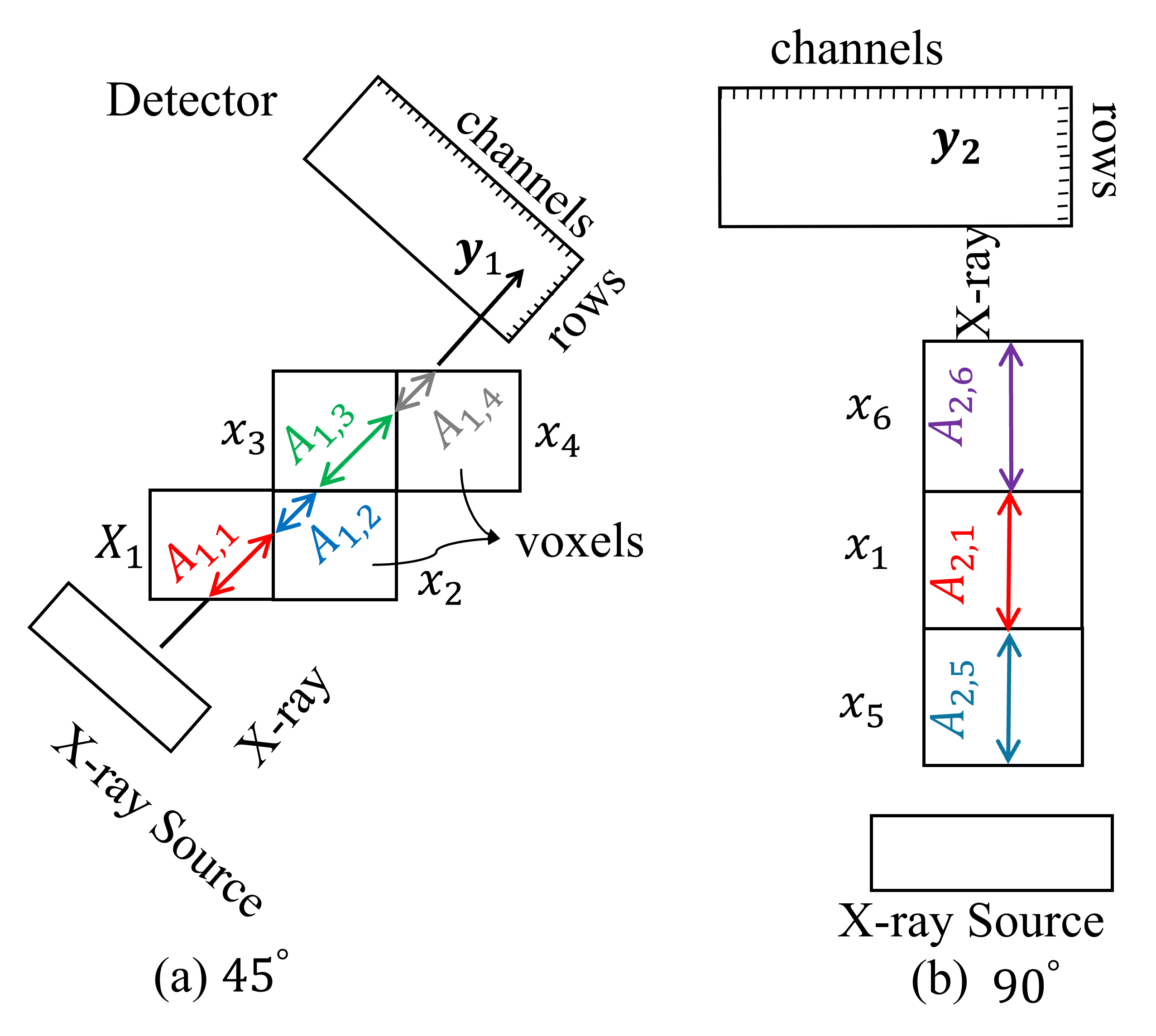}
\caption{Illustration of the CT system matrix. Each element $A_{i,j}$ represents the contribution of voxel $j$ to detector measurement $i$. For a distance-driven projector, this contribution is proportional to the intersection length between the X-ray and the voxel.}
\label{fig:A_matrix_cal}
\end{figure}

The accuracy and computational cost of iterative reconstruction depend heavily on how the system matrix is implemented. There are multiple ways of implementing the system matrix. The most accurate implementation is the \emph{distance-driven projector system matrix}, although it is also one of the most computationally and memory-intensive~\cite{de2004distance}. Figure~\ref{fig:A_matrix_cal} illustrates the construction of the system matrix. Consider the X-ray corresponding to detector measurement $Y_1$ in Figure~\ref{fig:A_matrix_cal}(a). At the current view angle, this X-ray intersects four voxels, denoted by $X_1$, $X_2$, $X_3$, and $X_4$. Since each measurement represents the accumulated attenuation along the X-ray path, the measurement $Y_1$ can be written as:

\begin{equation}
Y_1
=
A_{11}X_1
+
A_{12}X_2
+
A_{13}X_3
+
A_{14}X_4 \ ,
\end{equation}
where each coefficient $A_{i,j}$ is proportional to the length of intersection between voxel $j$ and the X-ray for detector measurement $i$. Figure~\ref{fig:A_matrix_cal}(b) illustrates a second detector measurement $Y_2$ acquired at a different projection view angle. The corresponding X-ray intersects a different set of voxels, $X_1$, $X_5$ and $X_6$, with a different set of lengths of intersections, therefore producing a different row of the system matrix.
Repeating this process for every X-ray across all view angles produces the complete linear system for the imaging forward operator: $
\mathbf{Y}
=
\mathbf{A}\mathbf{X}
+
\mathbf{E}
$ as in the general framework section~\ref{subsec:general_framework}.

Compared with the Radon transform, the distance-driven system matrix provides a much more accurate representation of the imaging system. Detector blur, detector sensitivity, focal spot size, geometric calibration, beam hardening, and other physical effects can all be incorporated into the system matrix by modifying the system matrix to account for the detector and physical effects, allowing iterative reconstruction to model the acquisition process much more faithfully than analytical FBP reconstruction.

Although this distance-driven system matrix projector closely matches the physical imaging process and generally produces the highest reconstruction accuracy, computing these geometric intersections is computationally expensive and requires significant memory storage.  For example, reconstructing a $512^3$ image volume from a $512^3$ sinogram would require a system matrix containing approximately $2^{54}$ elements, corresponding to tens of petabytes of storage in memory. Even with sparse matrix representations, since most system matrix elements are zeros, practical implementations of the distance-driven system matrix rarely store $\mathbf{A}$ explicitly. Instead, the required matrix elements are computed on demand during reconstruction.

To reduce computational cost, the majority of implementations for iterative reconstruction employ a simplified system matrix. A \emph{ray-driven projector} system matrix approximates each voxel as a point and computes the projection by interpolating image values along the X-ray path, avoiding explicit intersection-length calculations~\cite{galigekere2003cone}. Conversely, a \emph{voxel-driven projector} system matrix projects each voxel onto the detector and distributes its contribution to neighboring detector elements through interpolation~\cite{peters1981algorithms}. Although these approximate projectors introduce modest modeling errors compared with distance-driven implementation, they substantially reduce computational complexity and are widely used in practical reconstruction systems.

Regardless the system matrix is implemented as distance or voxel or ray-driven, the system matrix is too large to invert directly for a closed-form solution. Consequently, image reconstruction is formulated as an optimization problem whose solution provides the reconstructed image. Often, iterative reconstruction formulation assumes all measurements are equally reliable and seeks the image volume that minimizes the least-squares projection error, namely:

\begin{equation}
\hat{\mathbf{X}}
\gets 
\arg\min_{\mathbf{X\ge 0}}
\|
\mathbf{Y}
-
\mathbf{A}\mathbf{X}
\|^2 \ ,
\label{eq:ls}
\end{equation}
where $\|
\mathbf{Y}
-
\mathbf{A}\mathbf{X}
\|^2$ is short hand notation for $(\mathbf{Y}
-
\mathbf{A}\mathbf{X})^T (\mathbf{Y}
-
\mathbf{A}\mathbf{X})$ and $\mathbf{X}$ is strictly non-negative as all linear attenuation coefficients are non-negative.
To solve Equation~(\ref{eq:ls}) without inverting $\mathbf{A}$ directly, the above equation is solved numerically through iterative optimization, thereby earning the reconstruction approach the name ``iterative reconstruction". One common numerical method to solve the optimization problem is gradient descent. Differentiating the least-squares objective function gives:

\begin{equation}
\nabla_\mathbf{X}
=
-2
\mathbf{A}^{\dagger}
(
\mathbf{Y}
-
\mathbf{A}\mathbf{X}
) \ ,
\end{equation}
where $\nabla_\mathbf{X}$ represents the gradient for the least-squares objective function and $\mathbf{A}^{\dagger}$ denotes the adjoint of the system matrix. This gradient naturally gives rise to the computational steps of first computing $\mathbf{AX}$, which is also known as \emph{forward projection} for iterative reconstruction, and $\mathbf{A}^{\dagger}
(
\mathbf{Y}
-
\mathbf{A}\mathbf{X}
)$ as the \emph{Back-Projection} for iterative reconstruction.
These steps are repeated until the reconstruction converges. 

Note that gradient descent is not the only numerical method to solve the least square objective function. Different optimization algorithms differ primarily in how the image update is computed. Gradient descent, conjugate gradient, and quasi-Newton methods update the image using the gradient of the objective function, whereas coordinate-descent methods update one voxel or one subset of voxels at a time. Nevertheless, nearly all practical iterative reconstruction algorithms repeatedly evaluate the forward projection explicitly and its adjoint, back-projection, either explicitly or implicitly, making these two operations the dominant computational cost.

Compared with FBP, iterative reconstruction substantially improves image quality for sparse-view, limited-angle, and low-dose CT because it directly solves the discrete imaging model without relying on interpolation of the inverse Radon transform. Furthermore, the discrete system matrix provides a much more accurate representation of the imaging physics and can readily incorporate additional physical effects. However, these improvements come at a much higher computational cost. Modern iterative reconstruction requires repeatedly evaluating computationally intensive forward and back-projection operators over many iterations. Efficient implementations therefore rely heavily on GPU acceleration, distributed-memory parallelization, optimized memory access patterns, and other efficient computing techniques to achieve clinically practical reconstruction times~\cite{hidayetouglu2019memxct,9355274}.

The least-squares formulation in Equation~(\ref{eq:ls}) represents only the first generation of iterative reconstruction methods. In practice, different projection measurements exhibit different noise levels, and valuable prior knowledge about anatomical structures is often available. Incorporating these additional sources of information leads naturally to statistical iterative reconstruction, also known as model-based iterative reconstruction (MBIR), which is discussed in the next subsection.

\subsubsection{Statistical Iterative Reconstruction (Model-Based Iterative Reconstruction) for CT}
\label{subsubsec:mbir}

The iterative reconstruction methods described in the previous subsection formulate reconstruction as an ordinary least-squares optimization problem. Compared with analytical reconstruction, iterative reconstruction directly models the imaging system using a discrete system matrix rather than the continuous Radon transform. Consequently, it eliminates the interpolation errors associated with Filtered Backprojection, naturally accommodates more accurate models of imaging physics through the system matrix, and generally produces higher-quality reconstructions~\cite{wang2021physics}.

Despite these advantages, conventional iterative reconstruction still has two important limitations. First, the least-squares formulation in Equation~(\ref{eq:ls}) implicitly assumes that every measurement contributes equally to the reconstruction. In practice, however, different projection measurements exhibit different levels of statistical uncertainty. Measurements acquired with high photon counts are generally much more reliable than those acquired with low photon counts and therefore should contribute more strongly to the reconstruction. Second, the least-squares formulation relies solely on the acquired measurements and does not incorporate prior knowledge of the reconstructed image. Under sparse-view acquisition, limited-angle imaging, or low-dose CT, the reconstruction problem becomes increasingly ill-conditioned, allowing many possible image volumes to fit with the acquired sinogram. Prior knowledge about anatomical image structure can substantially reduce this uncertainty and improve reconstruction quality. Conventional iterative reconstruction, however, does not explicitly model either measurement uncertainty or prior image information.

Statistical Iterative Reconstruction, also known as \emph{Model-Based Iterative Reconstruction (MBIR)}, addresses these limitations by extending the least-squares formulation in Equation~(\ref{eq:ls}) to explicitly model both measurement uncertainty and prior knowledge of the reconstructed image~\cite{wang2021physics,wang2016high,Wang2017Massively3D,thibault2007three,bouman2022foundations}. The reconstruction is formulated as:

\begin{equation}
\hat{\mathbf X}
\gets
\arg\min_{\mathbf X\ge0}
\left\{
\|
\mathbf Y-\mathbf A\mathbf X
\|_{W}^{2}
+
\beta R(\mathbf X)
\right\},
\label{eq:mbir}
\end{equation}
where

\[
\|
\mathbf Y-\mathbf A\mathbf X
\|_{W}^{2}
=
(\mathbf Y-\mathbf A\mathbf X)^T
\mathbf W
(\mathbf Y-\mathbf A\mathbf X) \ .
\]

The first term $\|
\mathbf Y-\mathbf A\mathbf X
\|_{W}^{2}$ is called the \emph{data fidelity} (or data consistency) term, which encourages the reconstructed image to remain consistent with the measured projection data. The matrix $\mathbf W$ is the statistical weighting matrix for the measurements. The second term, $R(\mathbf X)$, is the image prior, which incorporates prior knowledge about the expected characteristics of the reconstructed image. The parameter $\beta$ controls the tradeoff between fitting the measured data and enforcing the image prior. 

Compared with conventional iterative reconstruction, the first major improvement introduced by MBIR is the statistical weighting matrix $\mathbf W$. Conventional iterative reconstruction assumes that every projection measurement is equally reliable and therefore contributes equally to the reconstruction. In practice, however, different X-ray paths receive different numbers of transmitted photons and consequently have different noise levels. Measurements with higher photon counts generally have lower statistical uncertainty and therefore provide more reliable information about the underlying image than measurements with fewer detected photons. By assigning larger weights to more reliable measurements and smaller weights to noisier measurements, MBIR allows the reconstruction algorithm to place greater emphasis on high-confidence projections while reducing the influence of noisy measurements. This statistical weighting becomes particularly important for low-dose CT, where the measurement noise varies substantially across detector channels.

The statistical weighting matrix is derived from the assumed noise model of the projection measurements. A common assumption is that the measurement vector follows a multivariate Gaussian distribution,
$\mathbf Y
\sim
\mathcal N
(
\mathbf A\mathbf X,
\mathbf\Sigma
)$,
where $
\mathbf\Sigma
=
\mathrm{diag}
(
\sigma_1^2,
\cdots,
\sigma_M^2
)
$ is the covariance matrix of the measurement noise, assuming that each measurement is independent~\cite{thibault2007three}. Under this assumption, the maximum-likelihood estimate leads to:
$
\mathbf W
=
\mathbf\Sigma^{-1}.
$. Consequently, each projection measurement is weighted inversely proportional to its noise variance. Measurements with smaller variance receive larger weights, whereas noisier measurements contribute less to the reconstruction.

For X-ray CT, the measurement variance is closely related to the transmitted photon statistics. Since projections with larger numbers of received photons exhibit lower statistical uncertainty, the diagonal elements of the weighting matrix are commonly approximated as:
$
\mathbf{W}_{ii}
\propto
\lambda_{t,i}
$,
where $\lambda_{t,i}$ denotes the number of transmitted X-ray photons for the $i^{th}$ measurement in the beer-lambert law in Equation~\eqref{eq:beer_lambert}. Consequently, projections with higher photon counts contribute more strongly to the reconstruction than noisier low-count measurements.

The second major improvement introduced by MBIR is the image prior $R(\mathbf X)$. Unlike the statistical weighting matrix, which models the reliability of the measurements, the image prior models our prior knowledge of the reconstructed image itself. Intuitively, the prior encodes prior knowledge about anatomical image structure and therefore constrains the set of feasible reconstruction solutions.
The importance of the prior becomes apparent when the projection data are incomplete or noisy. Under sparse-view, limited-angle, or low-dose acquisition, many different image volumes may produce projection data that fit the measurements nearly equally well. From the perspective of linear algebra, the reconstruction problem therefore becomes ill-conditioned and admits many possible solutions. The image prior reduces this ambiguity by favoring solutions that possess realistic anatomical characteristics, such as smooth homogeneous regions, sharp tissue boundaries, or sparse image representations. Consequently, the prior substantially improves reconstruction quality while suppressing noise and streak artifacts~\cite{bouman2022foundations}.

The choice of image prior has evolved considerably over the past three decades. Early MBIR methods employed quadratic ($L_2$) priors, which effectively suppress noise but tend to oversmooth anatomical edges. Subsequent methods introduced edge-preserving priors, including the $L_1$ norm, the Huber penalty, and generalized Gaussian Markov random field (QGGMRF) priors, which better preserve sharp boundaries while maintaining effective noise suppression~\cite{bouman2022foundations, tang2009performance}. More recently, deep learning has enabled learned priors through plug-and-play reconstruction, score-based diffusion models, and unrolled optimization networks~\cite{bouman2023generative, kamilov2023plug}. These deep learning priors capture substantially richer anatomical structures than handcrafted regularizers and represent one of the most active research directions in model-based image reconstruction.

\begin{figure}
\centering
\includegraphics[width=.9\linewidth]{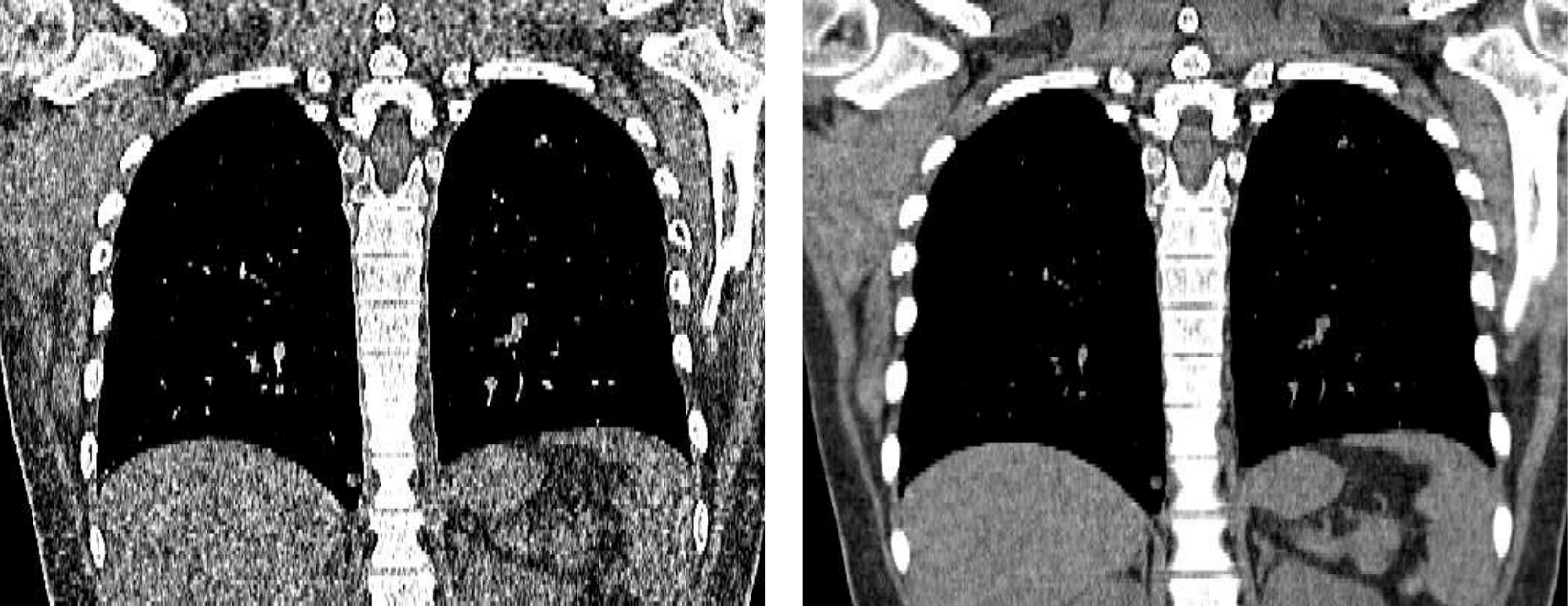}
\caption*{(a) Iterative Reconstruction \hspace{7em} (b) MBIR \hspace{8em}}
\caption{An example coronal-view image slice of a thoracic CT scan image reconstructed by the Siemens ADMIRE iterative reconstruction in soft tissue display window in (a), and  MBIR reconstruction for the same slice but with reduced image noise and artifacts in (b).}
\label{Fig:IR_vs_MBIR_1}
\end{figure}

\begin{figure}
\centering
\includegraphics[width=.9\linewidth]{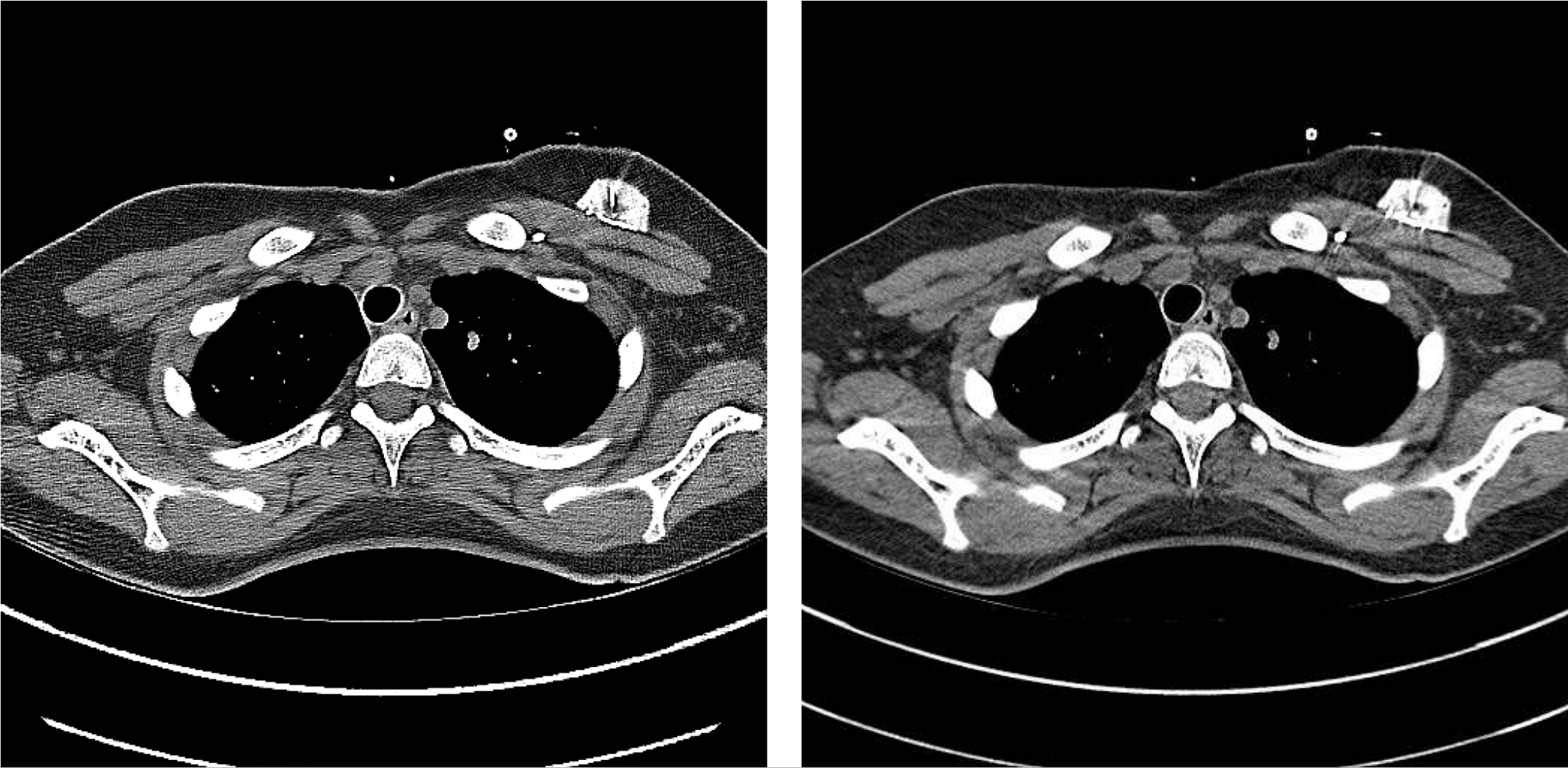}
\caption*{(a) Iterative Reconstruction \hspace{7em} (b) MBIR \hspace{8em}}
\caption{A thoracic CT scan with the Siemens ADMIRE iterative reconstruction shown on the left and MBIR reconstruction on the right.}
\label{Fig:IR_vs_MBIR_2}
\end{figure}

The practical impact of these improvements can be appreciated by comparing clinical reconstruction results. Figures~\ref{Fig:IR_vs_MBIR_1} and~\ref{Fig:IR_vs_MBIR_2} compare two representative thoracic CT slices reconstructed using a state-of-the-art commercial iterative reconstruction algorithm (Siemens ADMIRE) and MBIR algorithm. Although both methods substantially outperform analytical Filtered Backprojection, MBIR generally produces images with lower noise, reduced streak artifacts, improved edge preservation, and higher spatial resolution. These improvements arise from the combination of more accurate physical modeling, statistical noise modeling, and image priors discussed above. As a result, MBIR can often achieve higher image quality, particularly for low-dose and sparse-view CT acquisitions, although at the expense of substantially greater computational cost.

The MBIR formulation also has a natural Bayesian interpretation. From the perspective of Bayesian inference, the data fidelity term in Equation~(\ref{eq:mbir}) corresponds to the negative log-likelihood of the measured projection data, while the regularization term $R(\mathbf X)$ corresponds to the negative log-prior of the reconstructed image~\cite{thibault2007three}. Consequently, solving Equation~(\ref{eq:mbir}) is equivalent to computing the \emph{maximum a posteriori} (MAP) estimate of the unknown image. For simplicity, the medical imaging literature usually refers to $R(\mathbf X)$ simply as the \emph{prior}, even though mathematically it represents the negative logarithm of the prior probability distribution.

One important caveat is that the system matrix has different implementations as discussed in the previous subsection. The iterative reconstruction almost always uses a ray or voxel-driven projector system matrix implementation to save computations and reduce the expensive memory requirement. MBIR, however, typically employs distance-driven projectors because reconstruction quality is highly sensitive to inaccuracies in the forward model. The improved physical fidelity generally outweighs the additional computational cost, at the cost of much higher computing and memory requirements for computing the system matrix $\mathbf{A}$.

Like conventional iterative reconstruction, the MBIR optimization problem has no closed-form solution and must therefore be solved iteratively. Although numerous optimization algorithms have been proposed, they can be broadly divided into two categories: \emph{gradient-based optimization} and \emph{coordinate-descent optimization}. These methods solve the same optimization problem but exhibit different tradeoffs between convergence speed, memory requirements, and parallel computing efficiency.

Gradient-based optimization methods, including gradient descent, conjugate gradient, and quasi-Newton methods, update the entire image simultaneously during each iteration. Because every iteration operates on the complete image volume and projection dataset, these methods naturally expose abundant data parallelism and therefore map well onto multicore CPUs, GPUs, and distributed-memory HPC systems. Their primary disadvantage is relatively slow convergence. Practical CT reconstruction often requires hundreds or even thousands of iterations before convergence~\cite{wang2016high}. Various acceleration techniques, such as momentum methods, conjugate-gradient algorithms, quasi-Newton methods, and problem-specific preconditioners, can substantially improve convergence. However, many preconditioners are designed for specific imaging geometries or system matrices and therefore are not generally applicable across different reconstruction problems. Furthermore, each iteration repeatedly evaluates the complete forward and back-projection operators, resulting in substantial computational workload, memory traffic, and memory bandwidth requirements.

An alternative family of algorithms is \emph{coordinate-descent optimization}, also known as iterative coordinate descent (ICD) or Gauss--Seidel optimization~\cite{bouman1996unified}. Rather than updating all voxels simultaneously, coordinate-descent methods update one voxel (or a small block of voxels) at a time while keeping all remaining voxels fixed. Because only a small portion of the forward model is required for each update, these methods generally require less memory and less computation per update than gradient-based methods. They also tend to converge much more rapidly, often requiring only a few to a few tens of iterations to reach high-quality solutions. However, the sequential dependence between voxel updates makes coordinate-descent methods difficult to parallelize efficiently, limiting their scalability on modern GPU and distributed-memory architectures.

The choice between gradient-based optimization and coordinate-descent optimization therefore represents a fundamental tradeoff in efficient computing. Gradient-based methods provide excellent parallel scalability but typically require more iterations and greater computational resources, whereas coordinate-descent methods often converge much faster and are more memory efficient but expose much less parallelism. Regardless of the optimization algorithm, MBIR remains substantially more computationally demanding than analytical or conventional iterative reconstruction because it combines highly accurate forward models, statistical noise modeling, and sophisticated image priors. Consequently, practical MBIR implementations rely heavily on optimized projector algorithms, GPU acceleration, distributed-memory parallelization, and other HPC techniques to achieve clinically acceptable reconstruction times.

\subsubsection{Iterative and Statistical Iterative Reconstruction for MRI}
\label{subsec:mri_iterative}

Iterative and statistical iterative reconstruction for MRI follows essentially the same optimization framework as their CT counterparts. Iterative MRI employs the least-squares formulation in Equation~\eqref{eq:ls}, while statistical iterative MRI adopts the same model-based iterative reconstruction (MBIR) formulation in Equation~\eqref{eq:mbir}, which extends the least-squares objective by introducing a statistical weighting matrix $\mathbf{W}$ and an image prior $R(\mathbf{X})$~\cite{FesslerMRI,fessler2008image}. Consequently, the mathematical formulations, Bayesian interpretation, and numerical optimization algorithms discussed for CT apply almost directly to MRI.

The primary differences between CT and MRI lie in the implementation of the forward system matrix $\mathbf{A}$ and the formulation of the statistical weighting matrix $\mathbf{W}$, both of which are determined by the underlying imaging physics. In CT, $\mathbf{A}$ models X-ray projection through the patient, whereas in MRI it models Fourier encoding, receiver coil sensitivity, and k-space sampling during signal acquisition. Likewise, the weighting matrix reflects the statistical characteristics of the measurements. CT measurements are dominated by X-ray photon-counting statistics, whereas MRI measurements are dominated by thermal noise in the receiver electronics and are well approximated by a complex Gaussian distribution in k-space. The image prior $R(\mathbf{X})$, however, is largely independent of imaging modality and follows the same development as that described for CT, including quadratic, edge-preserving, and deep-learning priors.

For an ideal MRI system with a single receiver coil and fully sampled Cartesian acquisition, the forward operator reduces simply to the discrete Fourier transform:

\[
\mathbf{A}
=
\mathbf{F},
\]
where $\mathbf{F}$ denotes the discrete Fourier transform. This is the same forward model employed by analytical Fourier reconstruction. Consequently, unlike CT, iterative MRI does not introduce a different forward model. Instead, it solves the same Fourier imaging model through iterative optimization. This allows MRI to retain the exact physical imaging model while naturally incorporating additional constraints and prior information. Therefore, analytical Fourier reconstruction can be regarded as a special case of iterative MRI in which the forward operator is inverted directly rather than optimized numerically.

Modern MRI systems, however, rarely employ a single receiver coil. Instead, they use arrays of receiver coils for \emph{parallel imaging}, where each coil possesses a different spatial sensitivity profile and simultaneously acquires its own k-space measurements~\cite{SENSE,GRAPPA}. These complementary sensitivity profiles provide additional spatial encoding, allowing fewer k-space samples to be acquired while maintaining image quality. Consequently, the forward operator must account not only for Fourier encoding but also for coil sensitivity and k-space sampling. With multiple receiver coils, the MRI forward system matrix can therefore be expressed as:

\begin{equation}
\mathbf{A}
=
\mathbf{MFS} \ ,
\end{equation}
where $\mathbf{F}$ denotes the Fourier transform, $\mathbf{S}$ represents the coil sensitivity profile for all receiver coils, and $\mathbf{M}$ denotes the sampling mask that selects the acquired k-space locations based on the sampling trajectory. For fully sampled Cartesian MRI, $\mathbf M$ reduces to the identity operator. For accelerated MRI with subsampled k-space acquisition, it represents undersampled Cartesian acquisitions or more general non-Cartesian sampling trajectories such as radial or spiral imaging. Compared with analytical Fourier reconstruction, the additional operators $\mathbf S$ and $\mathbf M$ do not fundamentally change the imaging physics. Rather, they extend the Fourier model to account for practical MRI systems with multiple receiver coils and accelerated k-space acquisition. With the above formulation, the imaging forward model $\mathbf{Y=AX+E}$ can now be expressed as $\mathbf{Y=MFSX+E}$ for MR imaging. 

This formulation also provides an intuitive interpretation of parallel imaging. Rather than reconstructing each receiver coil independently, iterative reconstruction jointly estimates a single image that simultaneously fits the k-space measurements acquired by all receiver coils. The different coil sensitivity profiles provide complementary spatial information that compensates for missing k-space samples, thereby enabling accelerated MRI acquisition while maintaining high image quality. Consequently, parallel imaging enables accelerated MRI acquisition without requiring proportional increases in scan time.

The statistical weighting matrix $\mathbf{W}$ serves the same purpose as in CT by accounting for the statistical reliability of each measurement. The primary difference lies in the underlying noise model. Whereas CT measurements are dominated by X-ray photon-counting statistics, MRI measurements are dominated by thermal noise in the receiver electronics and are well approximated by a complex Gaussian distribution in k-space. Consequently, the weighting matrix $\mathbf{W}$ can be expressed as $\mathbf W=\mathbf\Sigma^{-1}$, where $\mathbf\Sigma$ is the noise covariance matrix of the receiver coils. In fact, many iterative reconstruction implementations for MRI treat different receiver coils as having independent noise with identical variance. Therefore, this covariance matrix $\mathbf\Sigma$ becomes the identity matrix, and the weighting matrix is often omitted in the implementation. However, a more accurate implementation for $\mathbf{W}$ needs to account for correlated noise among different receiver coils, allowing measurements with different statistical reliability to contribute appropriately to the reconstruction.

The image prior $R(\mathbf{X})$ is largely independent of imaging modality and therefore follows essentially the same evolution as described for CT MBIR. Early iterative MRI methods employed quadratic regularization, followed by edge-preserving priors such as the $L_1$ norm, total variation, Huber penalties, and generalized Gaussian Markov random field (QGGMRF) priors. More recently, deep learning priors based on plug-and-play methods, diffusion models, and unrolled optimization networks have become increasingly popular~\cite{yoo2021time, pour2019deep}. Although these reconstruction algorithms are mathematically identical to those used for CT, the learned prior must be trained using MR image datasets so that it captures the image contrast, anatomical structures, and statistical characteristics unique to MRI.

Unlike CT MBIR, the dominant computation in iterative MRI is repeated evaluation of the Fourier transform, coil-sensitivity operations, and k-space sampling, rather than computationally expensive geometric projection and back-projection. Consequently, iterative MRI generally requires substantially less computation per iteration than iterative CT reconstruction. Nevertheless, modern accelerated MRI using large receiver-coil arrays, highly undersampled or non-Cartesian k-space trajectories, compressed sensing, and learned deep-learning priors remains computationally demanding. Efficient implementations therefore continue to rely on optimized FFT algorithms, GPU acceleration, memory-efficient implementations, and distributed computing to achieve clinically practical reconstruction times.

\begin{figure}
\centering
\includegraphics[width=.7\linewidth]{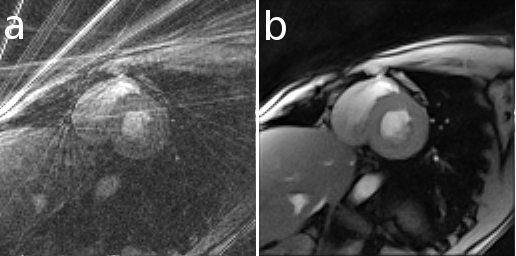}
\caption*{(a) Analytical \hspace{2em} (b) Statistical Iterative \hspace{3em}}
\caption{Comparison of analytical Fourier and statistical iterative reconstructions for accelerated cardiac MRI acquired in only 20 ms. Statistical iterative reconstruction substantially suppresses undersampling artifacts while preserving anatomical structures. Adapted from Uecker \emph{et al.}~\cite{uecker2010real} under the Creative Commons License.}
\label{Fig:MRI_analytical_iterative}
\end{figure}

To demonstrate the practical advantage of iterative MRI reconstruction, Figure~\ref{Fig:MRI_analytical_iterative} compares conventional analytical Fourier reconstruction with statistical iterative reconstruction for a cardiac MRI acquired in only 20~ms. The figure is reproduced from Uecker \emph{et al.}~\cite{uecker2010real} under the Creative Commons Attribution License. Such an aggressive acquisition time permits only sparse k-space sampling, causing direct analytical Fourier reconstruction to exhibit pronounced noise and undersampling artifacts. By solving the inverse problem using the MRI forward model together with statistical weighting and prior image information, statistical iterative reconstruction substantially suppresses these artifacts while preserving fine anatomical structures. This example illustrates why iterative and statistical reconstruction methods have become increasingly important for accelerated MRI, where reducing scan time without compromising image quality is a primary clinical objective.

\subsubsection{Iterative and Statistical Iterative Reconstruction for SPECT and PET}
\label{subsec:spect_pet_iterative}

Iterative and statistical iterative reconstruction for SPECT and PET follow the same mathematical framework introduced for CT and MRI. The imaging process is still described by the same general imaging forward model: $
\mathbf Y
=
\mathbf A\mathbf X
+
\mathbf E
$.
Consequently, SPECT and PET reconstruction remains an inverse problem that estimates the unknown activity distribution $\mathbf{X}$ from the measured photon counts $\mathbf{Y}$.

The physical meaning of this equation, however, differs from CT and MRI. In CT, $\mathbf X$ represents the attenuation coefficient distribution, whereas in SPECT and PET, $\mathbf{X}$ represents the radiotracer activity distribution. Likewise, the forward system matrix $\mathbf A$ no longer models X-ray transmission through the patient. Instead, it models how photons emitted from each voxel contribute to each detector measurement.

This forward imaging model is simply the discrete linear-algebra form of the continuous photon measurement equations introduced in Eqs.~\eqref{eq:spect_single_measurement} and~\eqref{eq:pet_single_measurement}. In these two equations, the expected photon count for a single measurement is determined by three components: the radiotracer activity distribution $\lambda_T(\mathbf r)$, the Beer--Lambert attenuation term, and the detector sensitivity $C_i(\mathbf r)$, where the detector sensitivity $C_i$ summarizes the probability that photons emitted from location r are detected. In practice, it accounts for detector response, detector blur, collimator response (for SPECT), detector normalization (for PET), and other scanner-specific physical effects. This sensitivity term is critical as emitted photons from a particular voxel have only a limited probability of reaching a given detector because they may be attenuated, scattered, or blocked before detection.

Iterative reconstruction discretizes the continuous activity distribution into image voxels, represented by the vector $\mathbf X$, and discretizes the detector measurements into the measurement vector $\mathbf Y$. The forward system matrix $\mathbf A$ is then obtained by discretizing the combined detector sensitivity and attenuation model, where each element of the system matrix, $A_{i,j}$, can be described as:

\begin{equation}
A_{i,j}
=
C_i(\mathbf r_j)
\exp
\left(
-
\int_{L_{ij}}
\mu(\mathbf r)\,d\mathbf r
\right),
\label{eq:A_matrix_spect_pet}
\end{equation}
where $\mathbf r_j$ denotes the location of voxel $j$. $L_{i,j}$ represents the ray path from the $j^{th}$ voxel to the $i^{th}$ measurement. Repeating this discretization for every detector measurement across all projection views produces the linear system:
$
\mathbf Y
=
\mathbf A\mathbf X
+
\mathbf E
$, which is simply the discrete counterpart of the continuous measurement equations introduced in the previous section.

This interpretation also explains why the SPECT and PET system matrix differs fundamentally from that of CT. In CT, the system matrix is determined primarily by imaging geometry through the intersection length between each X-ray and voxel. In SPECT and PET, the system matrix must model not only photon attenuation and intersection length, but also the probability that emitted photons are successfully detected through the detector sensitivity function $C_i(\mathbf r)$. Consequently, $C_i(\mathbf r)$ becomes a central component of the forward system matrix, accounting for detector response, collimator response (for SPECT), detector normalization (for PET), and other scanner-specific physical effects.

Compared with CT and MRI, the forward system matrix in SPECT and PET is therefore considerably more difficult to construct because it must accurately model both photon transport and detector physics. Accurately constructing the forward system matrix remains an active area of research because every additional physical effect improves image quality but also increases computational complexity. In practice, the detector sensitivity function $C_i(\mathbf r)$ is difficult to model precisely because it depends on scanner geometry, detector and collimator response (for SPECT), detector-pair normalization (for PET), finite detector resolution, and other hardware characteristics. Likewise, the measurement uncertainty $\mathbf E$ includes not only photon-counting statistics but also scattered photons, random coincidences (for PET), electronic noise, and calibration errors. Most practical reconstruction algorithms therefore employ approximations to both the forward system matrix and the measurement statistics in order to achieve computationally efficient reconstruction while maintaining clinically acceptable image quality.

Another important difference between SPECT/PET and CT lies in the statistical characteristics of the acquired measurements. In CT, the detected X-ray photons also follow a Poisson distribution. However, after applying the negative logarithm to the Beer--Lambert Law in Equation~\eqref{eq:X-ray-projection} to estimate the projections, the CT projections are well approximated as a Gaussian distribution given that the X-ray photons count is high. This approximation naturally leads to the weighted least-squares formulation introduced in the CT iterative reconstruction.

In SPECT and PET, however, the quantity of interest is the detected photon counts themselves rather than the logarithmically transformed projections. Consequently, detector measured photons $\mathbf Y$ are naturally modeled as a Poisson distribution vector~\cite{Cherry2012PhysicsNuclearMedicine, fessler2008image}:

\[
\mathbf Y
\sim
\mathrm{Poisson}\!\left(\mathbf A\mathbf X\right),
\]
where $\mathbf A\mathbf X$ denotes both the mean and variance for detected photon counts, namely $\mathbb{E}[\mathbf Y] = \operatorname{Var}(\mathbf Y)= \mathbf{AX}$. 
Because of this Poisson measurement model, the least-squares data fidelity term used in CT and MRI is replaced by the negative log-likelihood of the Poisson distribution for SPECT/PET iterative reconstruction:

\begin{equation}
\hat{\mathbf X}
=
\arg\min_{\mathbf X \ge 0}
\sum^M_{i=1}
\left\{
\mathbf A_{i,*}\mathbf X
-
Y_i
\log
(\mathbf A_{i,*}\mathbf X)
\right\} \ ,
\label{eq:spect_pet_ml}
\end{equation}
where $M$ is the number of measurements in sinogram. $\mathbf A_{i,*}$ represents the $i^{th}$ row of the system matrix.

Similar to CT and MRI, statistical iterative reconstruction incorporates prior knowledge of the reconstructed image, yielding the maximum a posteriori (MAP) estimation problem:

\begin{equation}
\hat{\mathbf X}
=
\arg\min_{\mathbf X \ge 0} \left\{
\sum^M_{i=1} \left(
\mathbf A_{i,*}\mathbf X
-
Y_i
\log
(\mathbf A_{i,*}\mathbf X)
\right)
+
\beta
R(\mathbf X)
\right\},
\label{eq:spect_pet_mbir}
\end{equation}
where $R(\mathbf X)$ denotes the image prior and $\beta$ controls the balance between measurement consistency and prior regularization. As in CT and MRI, the first term corresponds to the negative log-likelihood of the measurements, while the second term represents the negative log-prior. In addition, notice that, unlike CT and MRI, no explicit weighting matrix $\mathbf{W}$ appears in the data fidelity term. This is because the Poisson likelihood naturally accounts for the measurement variance through the statistical model itself, where the variance equals the expected photon count.

The numerical optimization algorithms are also closely related to those discussed for CT and MRI. Gradient-based optimization, conjugate-gradient methods, quasi-Newton methods, and iterative coordinate descent can all be applied to optimize the objective functions for iterative reconstruction and statistical iterative reconstruction for SPECT/PET. Although gradient-based optimization can be applied directly to the Poisson likelihood, nuclear medicine reconstruction has historically relied on a different optimization strategy. 

Historically, the dominant optimization strategy has been the \emph{Maximum-Likelihood Expectation-Maximization} (MLEM) algorithm~\cite{shepp1982maximum}. Rather than directly applying gradient descent to the Poisson log-likelihood, MLEM introduces the unknown origin of each detected photon as a latent variable and alternates between estimating the expected contribution of each voxel to the measured photons (Expectation step) and updating the activity distribution (Maximization step). This formulation naturally produces multiplicative image updates that preserve nonnegativity and monotonically increase the likelihood. The accelerated \emph{Ordered-Subsets Expectation-Maximization} (OSEM) algorithm further improves convergence by updating the image using subsets of the projection data and remains the dominant reconstruction algorithm in many commercial PET and SPECT systems~\cite{hudson1994accelerated}.

From the perspective of efficient computing, the iterative reconstruction families for SPECT and PET closely resemble those for CT. The dominant computational cost arises from repeatedly evaluating the forward projection and its adjoint while accurately modeling photon attenuation, detector response, scatter, and other physical effects. Consequently, efficient implementations rely heavily on optimized projector algorithms, GPU acceleration, memory-efficient implementations, and distributed computing to achieve clinically practical reconstruction times.

\subsection{Efficient Computing Considerations for Medical Image Reconstruction}
\label{sec:efficient_computing}

The previous subsections introduced the mathematical models and optimization algorithms underlying image reconstruction for CT, MRI, SPECT, and PET. Although these imaging modalities differ in their imaging physics, measurement statistics, and forward system matrix implementation, they all solve inverse problems governed by the same forward imaging model: $ \mathbf Y = \mathbf A\mathbf X + \mathbf E \ $, to estimate millions or even billions of unknown image voxels from noisy measurements.
Consequently, many computational challenges encountered during image reconstruction are remarkably similar across imaging modalities. Modern reconstruction algorithms are no longer limited primarily by mathematical formulation, but increasingly by computational efficiency. In practice, the difference between a high-quality reconstruction that finishes within seconds and one that requires hours is  determined as much by efficient computing algorithms as by imaging physics and numerical optimization.

Efficient image reconstruction therefore requires the co-design of medical physics, mathematical modeling, numerical optimization, computer architecture, and parallel computing. From a computational perspective, the major challenges can be organized into five interacting challenges: (1) Optimization efficiency; (2) Computational complexity; (2) Physics-aware forward system matrix modeling;
(4) Data movement and memory hierarchy;
(5) Clinical computing constraints.
These challenges are discussed in the following paragraphs.

\textbf{Optimization algorithms versus parallel scalability.}
The first computational challenge arises from the choice of the optimization algorithm itself. Optimization algorithms determine not only how quickly a reconstruction converges numerically but also how efficiently modern computing hardware can be utilized. Consequently, selecting an optimization algorithm involves balancing convergence rate, computational cost per iteration, and parallel scalability.

Gradient-based optimization methods, including gradient descent, conjugate-gradient methods, and quasi-Newton methods, update all image voxels simultaneously during each iteration. Because every voxel update can be computed simultaneously in gradient-based optimization, these algorithms map naturally onto multicore CPUs, GPUs, and SIMD architectures~\cite{wang2016high}. However, their numerical convergence is often relatively slow, particularly for ill-conditioned reconstruction problems, frequently requiring hundreds or even thousands of iterations before convergence~\cite{degirmenci2015acceleration, de2005study}. Various preconditioning techniques have therefore been developed to improve convergence, although effective preconditioners are usually highly problem dependent and often require careful algorithm-specific design~\cite{fessler1999conjugate}.

Iterative coordinate descent follows a very different strategy. Instead of updating all voxels simultaneously, it updates one voxel at a time while keeping all remaining voxels fixed~\cite{bouman1996unified}. Each update greedily minimizes the objective function with respect to a single voxel before proceeding to the next one. Because every update immediately incorporates the latest image estimate, coordinate descent often converges substantially faster than gradient-based methods, sometimes requiring only 3 to 6 iterations over the entire image volume~\cite{de2005study}. However, this sequential update strategy introduces strong data dependencies between neighboring voxel updates, making efficient parallelization considerably more difficult. Consequently, coordinate descent provides excellent numerical convergence but relatively poor parallel scalability~\cite{wang2016high}.

These observations illustrate an important tradeoff in efficient reconstruction algorithms. Algorithms with excellent numerical convergence are often difficult to parallelize, whereas algorithms that parallelize extremely well may require substantially more iterations before convergence. Therefore, the fastest-converging optimization algorithm is not necessarily the algorithm with the shortest wall-clock reconstruction time. This tradeoff between convergence rate and hardware utilization has motivated research on optimization algorithms designed specifically for modern parallel computing architectures.

Recent research has therefore explored optimization algorithms that balance numerical convergence with parallel scalability. One representative example is \emph{grouped coordinate descent}~\cite{fessler2002grouped} and its parallel implementation variants~\cite{Wang2017Massively3D}, which update a group of voxels simultaneously rather than updating a single voxel or the entire image. This approach forms a continuum between coordinate descent and gradient descent. If each group contains only one voxel, grouped coordinate descent reduces to conventional coordinate descent. Conversely, if all voxels belong to a single group, it becomes equivalent to gradient descent.

By updating multiple voxels simultaneously, grouped coordinate descent exposes significantly more parallelism while preserving much of the rapid convergence behavior of coordinate descent. Furthermore, vector instructions and GPU kernels can be efficiently applied to voxel updates within each group. However, simultaneous updates also violate the sequential dependency that contributes to the rapid convergence of conventional coordinate descent. As the group size increases, neighboring voxel updates increasingly interfere with one another, potentially slowing convergence or producing update overshoot~\cite{wang2016high}. Practical implementations therefore often introduce an under-relaxation (damping) parameter to stabilize the optimization while retaining high computational efficiency~\cite{wang2017high, zheng2000parallelizable}.

\textbf{Computational Complexity of Iterative Reconstruction.}
The second computational challenge is the computational complexity of the reconstruction algorithm itself. In general, improving reconstruction quality requires increasing the amount of computation performed during reconstruction.  From a computational perspective, image reconstruction methods form a hierarchy of increasing complexity. Analytical reconstruction methods, such as FBP for CT and inverse Fourier reconstruction for MRI, admit closed-form solutions and therefore require only a single reconstruction pass. Their computational cost is relatively low, enabling reconstruction within seconds on standard clinical workstations. This computational efficiency explains why analytical reconstruction remains the dominant reconstruction method in many routine clinical applications despite its relatively limited ability to model imaging physics and measurement statistics.

Iterative reconstruction substantially increases computational complexity because the forward system matrix and its adjoint must be evaluated repeatedly throughout the optimization process. Unlike analytical reconstruction, which computes the image only once, iterative reconstruction repeatedly performs forward projection, backprojection, and image updates until convergence. The computational cost therefore grows approximately linearly with the number of optimization iterations.
Statistical iterative reconstruction (MBIR) further increases computational complexity by incorporating more accurate physical imaging models, Bayesian image priors, and nonlinear optimization procedures. More sophisticated forward system matrix require more expensive projector evaluations, while complex image priors introduce additional computations during every optimization iteration.
More recently, AI-assisted reconstruction methods have introduced another level of computational complexity by incorporating deep neural networks into the image prior or directly into the reconstruction process. Although neural network inference can be relatively efficient once trained, training these models requires enormous computational resources and large annotated datasets.

The computational burden of statistical iterative reconstruction has historically limited its widespread clinical adoption. For example, when GE Healthcare first commercialized MBIR for clinical CT through the Veo reconstruction platform, reconstructing a single clinical scan typically required one to two hours using a dedicated computing cluster~\cite{mayo2014managing}. Although the resulting image quality exceeded that of conventional iterative reconstruction and FBP, the long reconstruction time substantially limited clinical workflow.

An important recent research direction seeks to combine the image quality of MBIR with the computational efficiency of analytical reconstruction. One representative approach is to train deep neural networks that learn the transformation from analytically reconstructed images to corresponding MBIR-quality images~\cite{ziabari20182}. Although training such models remains computationally expensive, inference requires only a single network evaluation and therefore approaches the computational cost of analytical reconstruction while producing image quality comparable to MBIR. However, ensuring the robustness, reliability, and generalizability of these AI learned reconstruction methods across diverse patient populations, scanner platforms, imaging protocols, and clinical conditions, remains an active area of research.

\textbf{Physics-aware Forward System Matrix Modeling.}
The third computational challenge arises from the construction of the forward system matrix. Unlike the optimization algorithm, which determines how the inverse problem is solved, the forward system matrix determines how faithfully the underlying imaging physics is represented. Because every additional physical effect modifies the forward operator, improvements in image quality are often accompanied by increased computational complexity. The forward system matrix therefore forms the primary interface between medical physics and efficient computing.

MRI provides perhaps the simplest forward operator among the four imaging modalities because the underlying physics is dominated by Fourier encoding. The forward operator is composed primarily of Fourier transforms, coil-sensitivity weighting, and sampling operations, all of which admit highly optimized numerical implementations. Consequently, evaluating the MRI forward operator has computational complexity comparable to analytical Fourier reconstruction and benefits directly from decades of FFT optimization.

In contrast, constructing accurate forward models for CT, SPECT, and PET is fundamentally different from MRI, and remains considerably more challenging. For CT, increasing physical fidelity requires modeling scanner geometry, detector response, focal spot blur, beam hardening, scatter, and photon attenuation. For SPECT and PET, the forward model becomes even more complicated because the detector sensitivity term $C_i(\mathbf r)$ must additionally account for collimator response (SPECT), detector normalization (PET), coincidence detection, attenuation correction, scatter correction, and other scanner-specific physical effects as mentioned in Section~\ref{subsec:spect_pet_iterative}. Incorporating additional physical models generally improves reconstruction accuracy but simultaneously increases computational complexity.

Memory storage presents another major challenge. Unlike analytical reconstruction, iterative reconstruction repeatedly evaluates the same forward and adjoint operators throughout optimization. Consequently, whether the system matrix is explicitly stored or computed on demand significantly affects the reconstruction times. For modern three-dimensional CT reconstruction, explicitly storing the complete system matrix may require hundreds of terabytes or even petabytes of memory depending on the image resolution and projector implementation. Consequently, most practical reconstruction algorithms compute system matrix elements on demand during forward and back projection rather than storing the matrix explicitly. Although this strategy dramatically reduces memory requirements, it substantially increases computational cost because the same system matrix entries must be recomputed repeatedly during every optimization iteration.

An alternative strategy is to compress the forward system matrix so that it can be stored explicitly while preserving sufficient physical accuracy. Such representations seek to trade a modest increase in storage for a substantial reduction in repeated projector computations. Therefore, developing compact system matrix representations remains an active research topic. If sufficiently accurate compressed representations could be stored entirely in memory, repeated projector computations could be largely eliminated, potentially reducing reconstruction time by orders of magnitude. Recent work on low-precision quantization of cone-beam CT system matrices demonstrates the feasibility of this direction~\cite{balke2018separable}. Extending these techniques to more challenging geometries, including helical multislice CT, SPECT, and PET, however, remains an open research problem.

\begin{figure}[t]
\centering
\includegraphics[width=.8\linewidth]{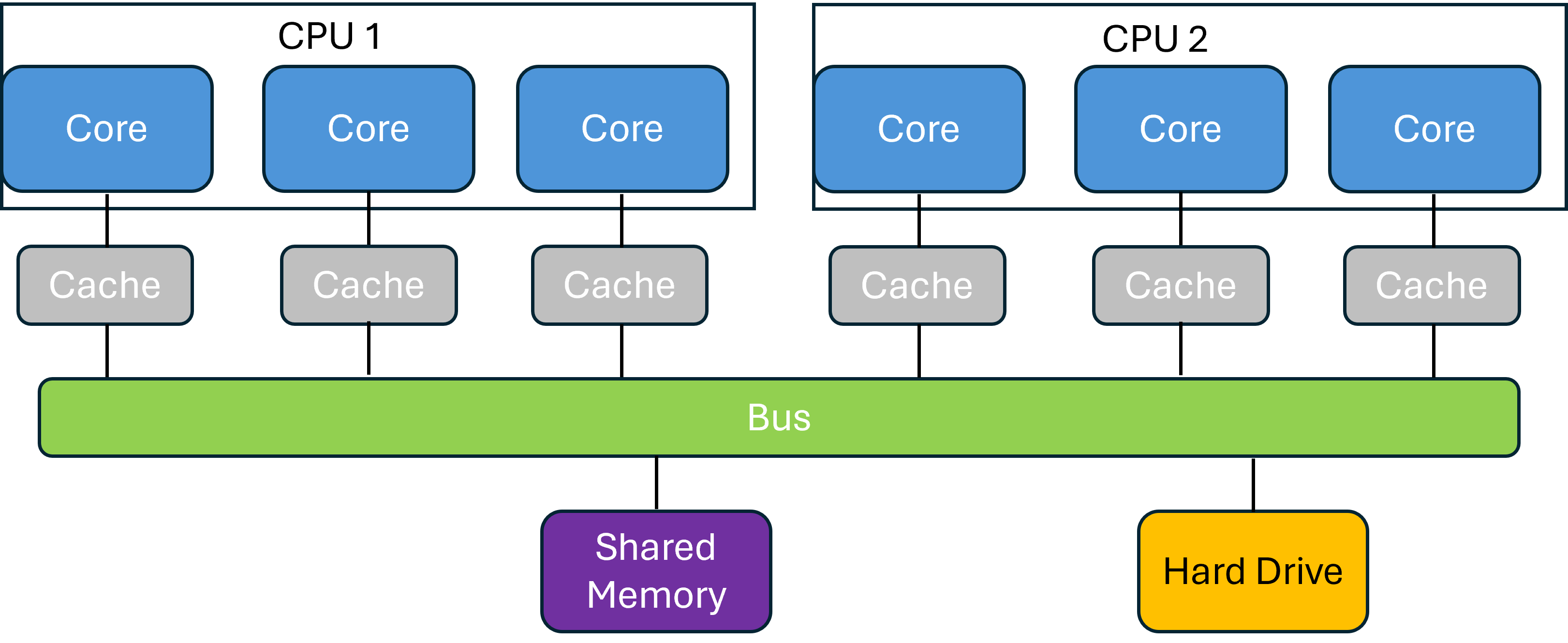}
\caption{An illustration of computer memory architecture, where data is accessed from IO device to memory, and then from memory to individual cache, and then from cache to cores for computing.}
\label{Fig:memory_architecture}
\end{figure}

\textbf{Data Movement and Memory Hierarchy.}
The fourth computational challenge is data movement through the memory hierarchy. For modern reconstruction algorithms, arithmetic computation is often no longer the dominant performance bottleneck. Instead, reconstruction speed is frequently limited by how efficiently measurement data and system matrix entries can be moved through the memory hierarchy. More specifically, modern processors can perform trillions of floating-point operations per second. However, these arithmetic units frequently remain idle because data cannot be supplied from the memory hierarchy fast enough to the computing units. Consequently, efficient image reconstruction has increasingly become a data-movement problem rather than a floating-point computing throughput issue.

Figure~\ref{Fig:memory_architecture} illustrates the hierarchical memory organization of a modern multi-core CPU system for example. During reconstruction, measurement data and system matrix entries must be accessed from the hard drive IO device into main memory, then through main memory to cache, before finally transferring the data from cache to each CPU core where arithmetic operations are performed. While floating-point operations typically require only a few processor cycles, accessing data from main memory may require hundreds of cycles, and accessing a secondary IO device is even slower. Consequently, data movement rather than arithmetic computation frequently dominates the overall reconstruction time.

\begin{figure}[t]
\centering
\includegraphics[width=.7\linewidth]{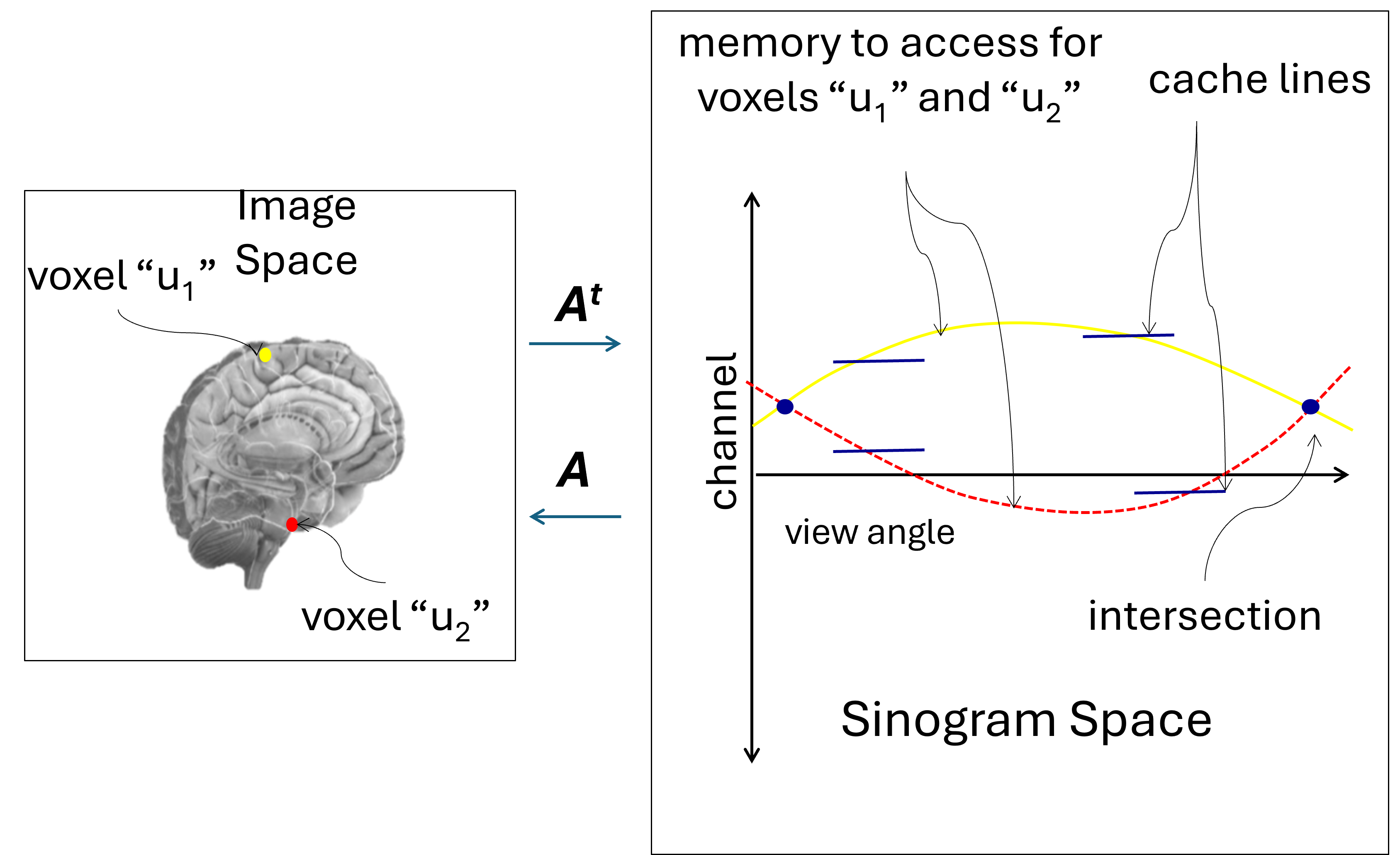}
\caption{Each voxel in the image space for CT, SPECT and PET corresponds to a sinusoidal trace in the sinogram. Different voxels correspond to different and unique traces in the sinogram.}
\label{Fig:voxel-trace}
\end{figure}

This challenge is particularly severe for iterative and statistical iterative reconstruction because the memory access pattern is dictated by the underlying imaging physics. In CT, SPECT, and PET, the measurements associated with a single image voxel correspond to a sinusoidal trace in the sinogram. Figure~\ref{Fig:voxel-trace} illustrates two example voxels and their corresponding sinusoidal traces in the sinogram. Since every voxel occupies a different spatial location, each voxel follows its own unique sinusoidal trace in the sinogram. Consequently, during image reconstruction, the algorithm must access measurements and their corresponding system matrix entries by following these physics-dependent sinusoidal traces in memory.

Unfortunately, this access pattern is poorly matched to modern computer architectures. CPUs and GPUs transfer data from main memory to cache using contiguous cache lines rather than individual measurements. As illustrated by the blue horizontal cache lines in Fig.~\ref{Fig:voxel-trace}, each cache line overlaps with only a small portion of the sinusoidal trace required for the current voxel update, while the remaining data loaded into the cache are not immediately used. Because each voxel has a distinct sinusoidal trace, these irregular memory accesses lead to poor cache locality, limited hardware prefetching, frequent cache misses, and significantly reduced effective memory bandwidth. Consequently, iterative reconstruction for CT, SPECT, and PET is often memory-bound rather than compute-bound.

\begin{figure}
\centering
\includegraphics[width=.7\linewidth]{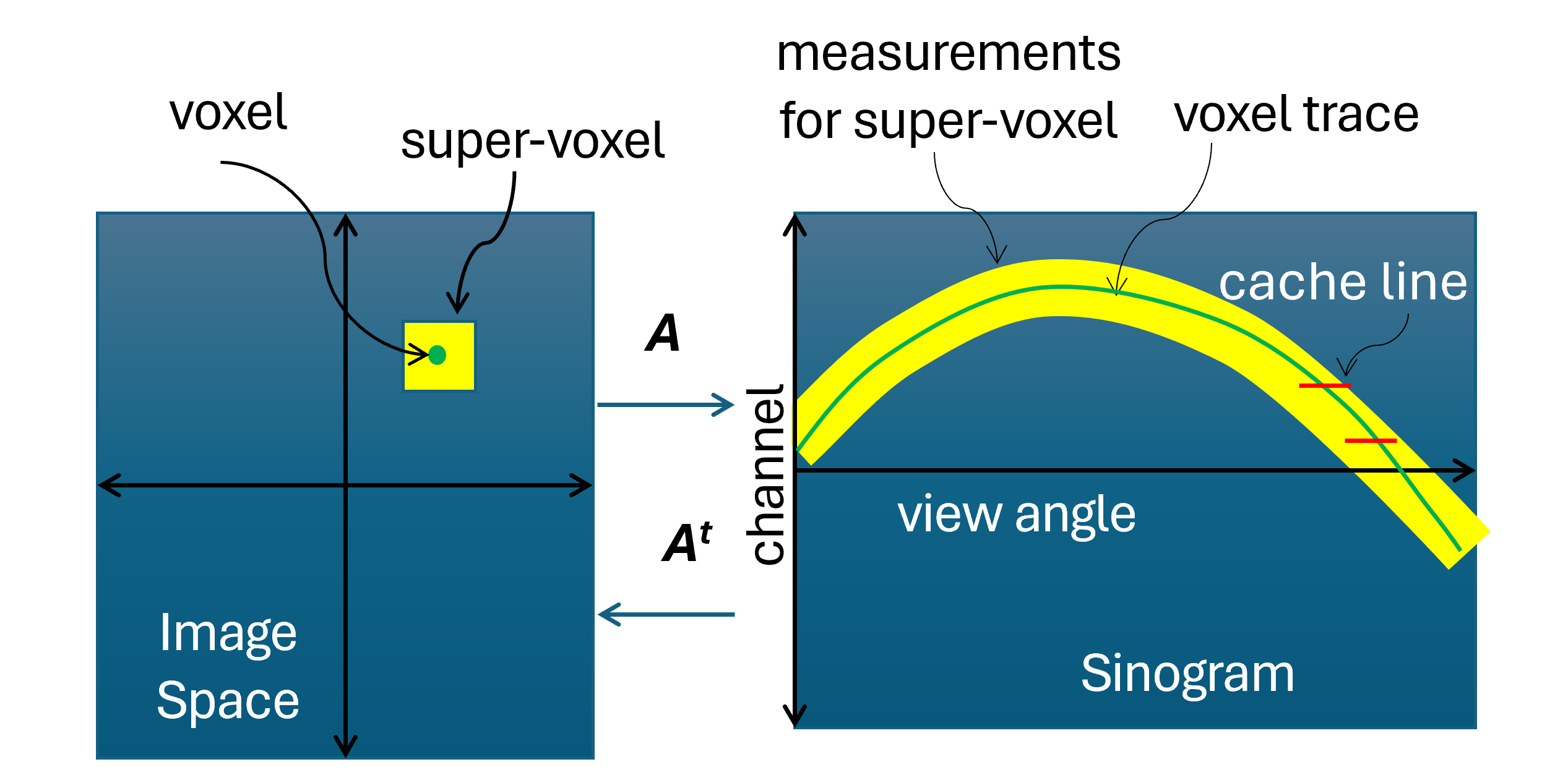}
\caption{Each super-voxel is a group of contiguous voxels in the shape of a square in the image space. The measurements for super-voxel correspond to a sinusoidal band in the sinogram, and each voxel within the super-voxel corresponds to a trace within the band in the sinogram.}
\label{Fig:supervoxel}
\end{figure}

To address this challenge, cache-aware reconstruction algorithms reorganize the computation according to both the imaging physics and the computer memory hierarchy. One representative example is the super-voxel framework~\cite{wang2016high}, illustrated in Fig.~\ref{Fig:supervoxel}. A super-voxel is defined as a square group of spatially neighboring voxels in the image space. For CT, SPECT, and PET, the measurements corresponding to all voxels within a super-voxel form a sinusoidal band in the sinogram, illustrated as the yellow band in Fig.~\ref{Fig:supervoxel}. Each voxel inside the super-voxel corresponds to its own sinusoidal trace within this band. During reconstruction, the measurements required for each voxel are still accessed by following its sinusoidal trace. However, data are transferred from main memory to cache in contiguous cache lines, illustrated by the red dashed lines in Fig.~\ref{Fig:supervoxel}.

Although each cache line overlaps only partially with the sinusoidal trace of the current voxel, the remaining measurements loaded into the cache are not wasted. Instead, they still lie within the same sinusoidal band and will be accessed from cache, instead of memory, shortly afterward when updating neighboring voxels in the same super-voxel. Therefore, the super-voxel algorithm completes the updates for all voxels within a super-voxel before moving to the next one, allowing measurements already loaded into the cache to be reused repeatedly rather than fetched again from main memory or secondary storage. In the example shown in Fig.~\ref{Fig:supervoxel}, the red dashed cache lines overlap only partially with the green voxel's sinusoidal trace, yet the remaining cached measurements still belong to the yellow sinusoidal band and are subsequently reused by neighboring voxels within the same super-voxel. This significantly improves cache utilization, reduces memory traffic, and accelerates image reconstruction. Compared with conventional reconstruction with the super-voxels, the super-voxel framework achieves approximately a 10-fold speedup on a single CPU core and an 187-fold speedup on a 20-core shared-memory system~\cite{wang2016high}.

In fact, this example illustrates a broader principle that extends beyond CT reconstruction. Efficient image reconstruction is not achieved solely through faster processors or additional computing hardware. Instead, high-performance reconstruction requires co-design across medical physics, mathematical modeling, numerical optimization, computer architecture, and parallel computing. As image reconstruction algorithms continue to incorporate increasingly sophisticated physical models and large AI foundation models, such interdisciplinary algorithm--hardware co-design will likely become one of the central research directions in computational imaging.

\textbf{Clinical Computing Constraints.}
The final challenge is that medical image reconstruction operates under computing constraints fundamentally different from those encountered in traditional HPC. Unlike many large-scale scientific computing applications that execute on leadership-class supercomputers, medical image reconstruction is typically performed directly on the clinical imaging scanner or on an attached workstation. Consequently, reconstruction algorithms operate under much tighter computational constraints, including limited processor count, memory capacity, power consumption, and reconstruction latency.

These practical constraints fundamentally distinguish medical image reconstruction from many traditional HPC applications. Rather than maximizing scalability across thousands of processors, reconstruction algorithms must maximize computational efficiency on relatively modest computing platforms while satisfying strict clinical turnaround times. Consequently, algorithm design must simultaneously consider numerical convergence, memory efficiency, parallel scalability, communication overhead, hardware utilization, and clinical workflow.

\textbf{Conclusion.} The five computational challenges discussed above are tightly coupled. Optimization algorithms determine convergence behavior; reconstruction algorithms determine computational complexity; imaging physics determines the forward system matrix; the forward model dictates memory access patterns and data movement; and clinical constraints ultimately determine the available computing resources. Consequently, efficient image reconstruction cannot be achieved by improving only one component of the pipeline.

Looking forward, image reconstruction will increasingly combine sophisticated physics-based forward models with large AI foundation models. These advances will dramatically increase computational demand while simultaneously raising expectations for image quality and clinical turnaround time. Meeting these competing objectives will require continued co-design across medical physics, applied mathematics, artificial intelligence, computer architecture, and high-performance computing. Such interdisciplinary integration is likely to define the next generation of computational imaging systems.

%% file: acknowledgement.tex
%
%

\extrachap{Acknowledgements}
This manuscript has been authored by UT-Battelle, LLC under Contract No. DE-AC05-00OR22725 with the U.S. Department of Energy. The United States Government retains and the publisher, by accepting the article for publication, acknowledges that the United States Government retains a non-exclusive, paid-up, irrevocable, world-wide license to publish or reproduce the published form of this manuscript, or allow others to do so, for United States Government purposes. The Department of Energy will provide public access to these results of federally sponsored research in accordance with the DOE Public Access Plan (\url{http://energy.gov/downloads/doe-public-access-plan}).